# Sharing information across patient subgroups to draw conclusions from sparse treatment networks

**Theodoros Evrenoglou[1], Silvia Metelli[1], Johannes-Schneider Thomas[2], Spyridon Siafis[2], Rebecca M. Turner[3], Stefan Leucht[2], Anna Chaimani[1]**

[1] *Université Paris Cité, Research Center of Epidemiology and Statistics (CRESS-U1153), INSERM, Paris, France*

[2] *Department of Psychiatry and Psychotherapy, School of Medicine, Technical University of Munich, Germany*

[3] *MRC Clinical Trials Unit, University College London, London, UK*

*Corresponding Author:

Theodoros Evrenoglou

Université Paris Cité, Research Center of Epidemiology and Statistics (CRESS-U1153), INSERM, Paris, France

Hôpital Hôtel-Dieu, 1 Place du Parvis Notre-Dame, 75004 Paris

email: tevrenoglou@gmail.com

**Abstract:** Network meta-analysis (NMA) usually provides estimates of the relative effects with the highest possible precision. However, sparse networks with few available studies and limited direct evidence can arise, threatening the robustness and reliability of NMA estimates. In these cases, the limited amount of available information can hamper the formal evaluation of the underlying NMA assumptions of transitivity and consistency. In addition, NMA estimates from sparse networks are expected to be imprecise and possibly biased as they rely on large sample approximations which are invalid in the absence of sufficient data. We propose a Bayesian framework that allows sharing of information between two networks that pertain to different population subgroups. Specifically, we use the results from a subgroup with a lot of direct evidence (a dense network) to construct informative priors for the relative effects in the target subgroup (a sparse network). This is a two-stage approach where at the first stage we extrapolate the results of the dense network to those expected from the sparse network. This takes place by using a modified hierarchical NMA model where we add a location parameter that shifts the distribution of the relative effects to make them applicable to the target population. At the second stage, these extrapolated results are used as prior information for the sparse network. We illustrate our approach through a motivating example of psychiatric patients. Our approach results in more precise and robust estimates of the relative effects and can adequately inform clinical practice in presence of sparse networks.

# 1. Introduction

Network meta-analysis (NMA) has become an essential tool for the comparative effectiveness research of healthcare interventions since it allows integrating all available information on a specific disease and usually provides estimates of relative effects with the highest possible precision[1,2]. However, networks of interventions with limited data that fail to provide useful conclusions may arise under certain situations. A recent empirical study found 92 (7.4%) published NMAs that included more treatments than the number of studies in a sample of 1236 NMAs of randomized controlled trials (RCTs) with at least four interventions[3]. These numbers suggest that the phenomenon of 'sparse' networks, namely networks with limited direct evidence in the form of few direct comparisons and/or few - 1 or 2 - studies per comparison, is not very rare in the literature. For example, this is often the case for 'sensitive' subgroups of the population that cannot be easily included in trials, such as children[4], elder patients, or individuals with multimorbidity.

Results from such networks of interventions are accompanied with substantial uncertainty not only in the estimation of the relative treatment effects but also in the plausibility of the underlying assumptions since their formal evaluation is impossible[5]. In addition, the large sample approximations on which NMA models typically rely upon fail in the presence of only a handful of studies per comparison. Hence, lack of robustness and potentially limited reliability are common issues that might be encountered when analyzing sparse treatment networks[6]. To avoid these issues, it might seem reasonable to wait until more studies become available for the outcome(s) of interest. However, the pace of trial production in these contexts is usually very slow and the chance to obtain new study results shortly is low.

The use of external evidence in the form of informative prior distributions has been suggested previously in meta-analysis, either to achieve more accurate estimation of the heterogeneity parameter[7,8] or for the incorporation of non-randomized evidence[9]. Here, we introduce a new framework that allows sharing information between two networks of treatments that pertain to different subgroups of the population. Our framework refers to the case when only aggregate data are available; in the presence of individual participant data, other approaches may be used, such as population adjustment methods[10]. Specifically, we use the results from a subgroup with a lot of direct evidence (i.e. many direct comparisons and many studies per comparison) forming a 'dense' network to construct informative priors for the relative effects in the target subgroup forming a sparse network. This is a two-stage approach where at the first stage we synthesize the data from the dense network using a hierarchical NMA model. This includes a location parameter that shifts the distribution of the relative effects to make them applicable to the population of the sparse network. We also add a scale parameter that allows us to further downweight the external data and to reduce their influence. At the second stage, the results from the first stage are used as prior information in the analysis of the sparse network. We inform the location and scale parameters either through the data or using expert opinion. We illustrate our approach through an example examining the relative effects of antipsychotic treatments in two subgroups of patients: a target subgroup of children-adolescents (CA) that forms a rather sparse network and a subgroup of chronic adult patients with acute exacerbation of schizophrenia, referred here as “general patients” (GP), that forms a dense network.

# 2. Motivating dataset

Our work is motivated by a recent article investigating the differences in the treatment effects among several subgroups of schizophrenia patients[11]. The aim of the article was to compare the subgroup of GP, defined as chronic patients with an acute exacerbation of positive symptoms, with more specific subgroups of patients: CA, first-episode patients, treatment-resistant patients, patients with negative symptoms, patients under substance abuse, and elder patients. In contrast to the dataset for GP, all these specific subgroups were informed by very limited direct evidence and therefore, the original authors, who acknowledged the problems of analyzing sparse networks, only used pairwise meta-analysis. To illustrate our approach, we use the data from two of the above subgroups: CA and GP. The former is considered the target subgroup which forms a sparse network and the latter is the only available subgroup with abundant direct evidence (forming a dense network) that can be safely used to elicit informative priors for CA.

The CA network comprises 19 randomized controlled trials (RCTs) published between 1973 and 2017 comparing 14 antipsychotics and placebo, and providing direct information for 21 out of 105 possible comparisons (**Figure 1a**). Out of these 21 comparisons, two include 2 RCTs and the rest include 1 RCT only. In addition, many of these RCTs are small, with a median study sample size of 113 patients. The pace of new RCT production for this subgroup is very slow (on average one RCT every 2.3 years), which confirms the difficulty of conducting RCTs for this group of patients. On the other hand, the network for general patients contains 255 RCTs comparing 33 antipsychotics and placebo, and forming 116 direct comparisons (**Figure 1b**). The two networks are distinct, which means that they do not have any RCT in common. However, some drugs are included in both networks; this is a necessary condition to allow sharing information between them. The outcome of interest is the reduction of the overall schizophrenia symptoms which is measured in a continuous scale. Following the original article, we used the standardized mean difference (SMD) as effect measure since the different RCTs have used different scales.

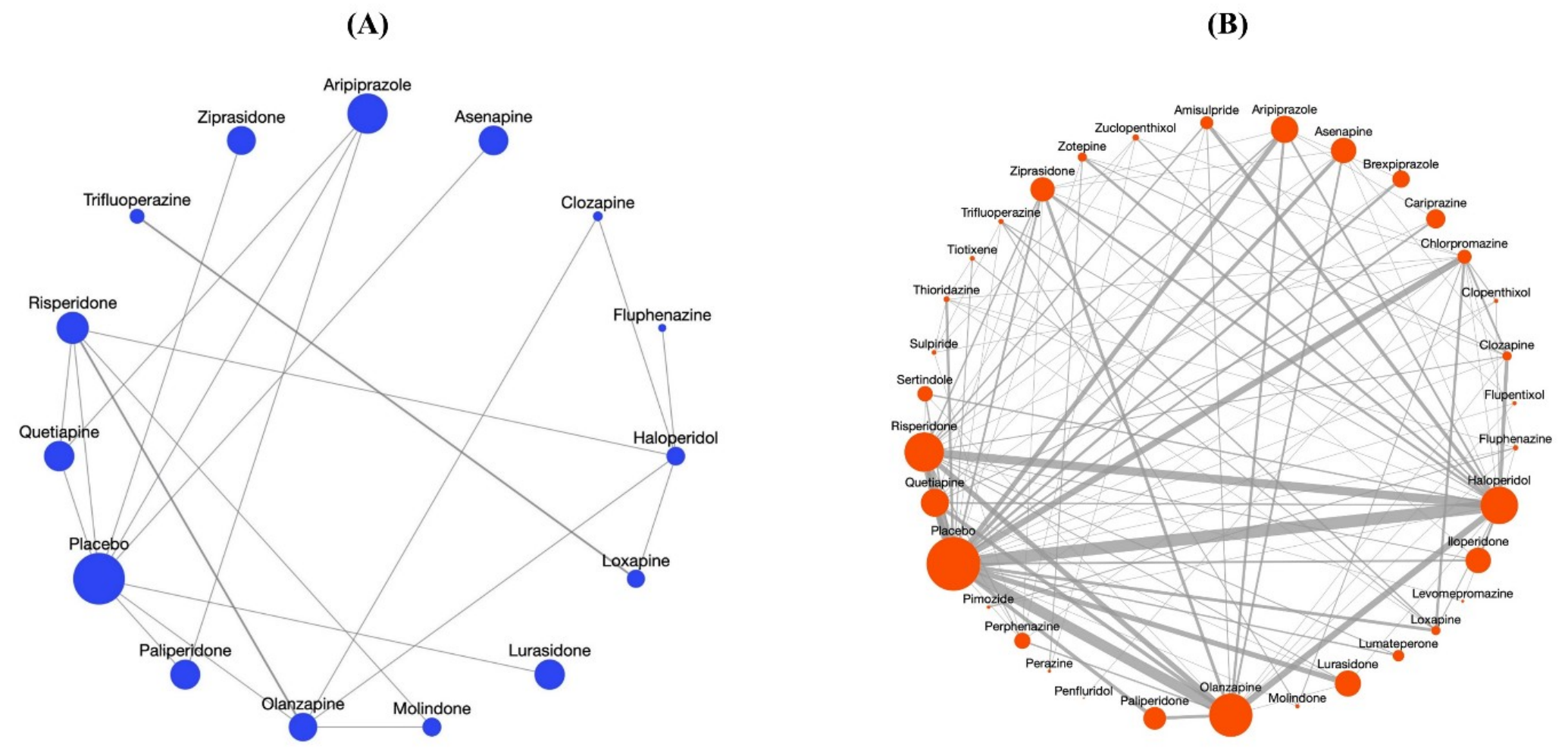

**Figure 1:** Sparse network for the population of CA (panel A) and informative network for GP (panel B). The size of the edges (lines in the networks) is proportional to the number of studies that compare the respective treatments while the size of the nodes is proportional to number of patients that received each treatment.

# 3. Methods

## 3.1 Notation and general setting

Suppose we have a set of $N = n_1 + n_2$ studies. Studies $\{1,2,\dots,n_1\}$ inform the target population subgroup $P_1$ and form a sparse network while the rest $\{n_1+1, n_1+2,\dots,N\}$ inform a different population subgroup $P_2$ and form a dense network. Suppose also that $T_1$ treatments have been evaluated for population $P_1$ and $T_2$ treatments for population $P_2$. We denote with $T_a = T_1 \cup T_2$ the set of all available treatments and with $T_c = T_1 \cap T_2$ the set of common treatments included in both networks; the latter is the group of treatments for which information can be extrapolated from $P_2$ to the target subgroup $P_1$. Each study $i$ provides the observed mean (change score/endpoint) $y_{it_k}$ for the treatment $t_k \in T_a$ of arm $k$ with $k = 1, \dots, K_i$ and $K_i$ the total number of arms in $i$. For simplicity, in what follows we write the $y_{it_k}$ as $y_{ik}$, its variance $sd_{ik}^2$, and the respective 'true' mean is denoted as $\theta_{ik}$. We consider treatment $j = 1$ as the common network reference for both $P_1$ and $P_2$ and the relative effects $\mu_{1j}$ ($j > 1, j \in T_a$) as the basic parameters. For every study, we arbitrarily choose the treatment of arm $k = 1$ ($t_1$) as the study-specific baseline treatment.

### 3.2 Models for sharing information between two population subgroups

Different models stem from different assumptions about the relation between the two subgroups $P_1$ and $P_2$. We start with the description of the standard hierarchical NMA model which in the present setting synthesizes the two subgroups in a naïve way, namely as if they were equivalent. Then, we move to more plausible NMA models in which we incorporate a location parameter acknowledging the difference between $P_1$ and $P_2$ and a scale parameter aiming to increase the uncertainty of the studies in subgroup $P_2$.

#### 3.2.1 Naïve synthesis

In this approach the two subgroups are combined together as if all participants were coming from the same population $P$. In other words, this approach makes the strong assumption that $P_1$ and $P_2$ are equivalent, namely $P_1 \equiv P_2 \equiv P$. Hence, for every study $i = 1, \dots N$ and treatment $t_k \in T_\alpha$, the observed means follow a normal distribution

$$y_{ik} \sim \mathrm{N}(\theta_{ik}, sd_{ik}^2). \qquad (1)$$

Then, the underlying SMD between every treatment $t_2, t_3, \dots, t_{K_i}$ versus the baseline treatment $t_1$ is

$$\delta_{i,1k} = \frac{\theta_{ik} - \theta_{i1}}{sd_i^{pooled}}, \qquad (k > 1), \qquad (2)$$

with $sd_i^{pooled} = \sqrt{\sum_{k=1}^{K_i} \frac{n_{ik} sd_{ik}^2}{n_{ik} - K_i}}$ the between-arm standard deviation. Under the random-effects assumption, the underlying relative effects are assumed to follow a multivariate normal distribution

$$\boldsymbol{\delta}_{\mathbf{i}} \sim \mathrm{N}_{K_i - 1}(\boldsymbol{\mu}, \boldsymbol{\Sigma}), \tag{3}$$

where $\boldsymbol{\mu} = (\mu_{12}, \dots, \mu_{1K_i})$ is the vector of the summary relative effects and $\boldsymbol{\Sigma}$ the between-study variance-covariance matrix with entries $\tau^2$ (the common heterogeneity variance across comparisons) in the diagonal and $\frac{\tau^2}{2}$ in the off-diagonal. Finally, the transitivity assumption for every pair of treatments $j, l \in T_a$ $(j, l > 1)$ implies that

$$\mu_{jl} = \mu_{1l} - \mu_{1j}.$$

Non-informative prior distributions are usually employed for every $\mu_{1j}$ and $\tau$, such as $\mathrm{N}(0{,}10000)$ and the half-normal $\mathrm{HN}(1)$ respectively.

Apart from the strong assumption about the similarity of the two subgroups, a further problem in the present setting is that the final results for the joint population $P$ will be dominated by the dense network ($P_2$) rather than from the target subgroup $P_1$. The same would apply if the results of the dense network were used directly as prior information for $P_1$. To avoid this issue, in the proposed approach we extrapolate the results from $P_2$ to $P_1$, acknowledging differences between the two subgroups, before forming informative prior distributions for $P_1$.

#### 3.2.2 Using the external subgroup $P_2$ to construct informative priors for the target subgroup $P_1$

This is a two-stage approach where at the first stage we extrapolate the results from the dense network of subgroup $P_2$ to $P_1$ and at the second stage we use predictions from this extrapolation to form prior distributions for the analysis of $P_1$.

*First stage: Extrapolating external results to the target population subgroup*

In contrast to the naïve synthesis, the main assumption here is that the two subgroups have some underlying differences in the treatment effects which can be considered as a different location of their outcome distributions. To incorporate some uncertainty about this assumption, in Equation (*1*) we introduce a scale parameter $w_i \in (0,1]$ that inflates the variance of each $y_{ik}$ in subgroup $P_2$ for each $i = n_1 + 1, n_1 + 2, \dots, N$. This modifies Equation (1) into

$$y_{ik} \sim \mathrm{N}\left(\theta_{ik}, \frac{sd_{ik}^2}{w_i}\right). \tag{4}$$

Equation (4) implies that the parameters $w_i$ only affect the variance of the the mean scores $y_{ik}$ and not the true study-specific SMDs which are still the same as in Equation (2). The location parameters $\boldsymbol{\beta}$ that aim to shift the original distribution of the SMDs in $P_2$ towards the distribution of $P_1$ are then added in Equation (3):

$$\boldsymbol{\delta}_\mathbf{i} \sim \mathrm{N}_{\mathrm{K_i}-1}(\boldsymbol{\mu}^{P_2} - \boldsymbol{\beta}, \boldsymbol{\Sigma}), \tag{5}$$

with $\boldsymbol{\beta} = \left(\beta_{12}, \dots, \beta_{1K_i}\right)$ being the vector of the 'true' differences between the comparison-specific SMDs of the two subgroups. The extrapolated means $\boldsymbol{\mu}^* = \boldsymbol{\mu}^{P_2} - \boldsymbol{\beta}$ for population $P_2$ in Equation (5) are then considered similar to those expected for population $P_1$.
By fitting this modified NMA model we obtain the extrapolated SMD estimates $\hat{\mu}_{jl}^*$ for every pair of treatments $j, l \in T_c$. Subsequently, the predictive distributions[12] of $\mu_{jl}^{new}$ namely $\mathrm{N}\left(\mu_{jl}^*, (\tau^{P_2})^2\right)$ are used at the second stage as prior distributions for $P_1$. Note that the parameter $(\tau^{P_2})^2$ represents the heterogeneity across the studies in $P_2$.

*Second stage: NMA of the target subgroup using informative prior distributions for relative effects*

Here the model of Section 3.2.1 is used for $i = 1, \dots, n_1$. The key difference is that for the $\mu_{1l}$s we use as prior distribution $\mu_{1l}^{P_1} \sim \mathrm{N}(\hat{\mu}_{1l}^{new}, var(\hat{\mu}_{1l}^{new}))$ where $\hat{\mu}_{1l}^{new}$ and $var(\hat{\mu}_{1l}^{new})$ are the posterior mean and variance of $\mu_{1l}^{new}$, respectively. The use of informative priors is expected to improve the precision in the final NMA estimates without dominating the analysis. This is because the priors are constructed at the first stage by extrapolating the results of the dense network to those expected for the target sparse network.

### 3.2.3 Informing the location and the scale parameters

*Constructing prior distributions for $\beta$s using the data*

To obtain an estimate of the difference in the outcome distributions between the two subgroups, we first use the estimates $u_{1j}^{P_1}$ and $u_{1j}^{P_2}$ obtained from separate pairwise meta-analyses for each population and each pair of treatments $\{1, j\}, j \in T_c$. To allow the estimation of heterogeneity for every comparison in $P_1$, here we assume a common comparison-specific heterogeneity across the two subgroups $\left(\sigma_{1j}^{P_1}\right)^2 = \left(\sigma_{1j}^{P_2}\right)^2 = \sigma_{1j}^2$. Using the difference of these estimates $d_{1j} = u_{1j}^{P_2} - u_{1j}^{P_1}$, we construct prior distributions for the parameters $\beta_{1j}$, namely

$$\beta_{1j} \sim \mathrm{N}\left(d_{1j}, var\left(d_{1j}\right)\right). \tag{6}$$

For treatment comparisons not evaluated in $P_2$ we use a non-informative normal prior $\mathrm{N}(0{,}10000)$.

*Constructing prior distributions for $\beta$s using expert opinion*

Elicitation of expert opinion can be undertaken with several methods such as face to face interviews, software tools, or questionnaires. The pooled change score of each treatment $j \in T_c$ in $P_2$ can be used as a reference for the experts who need to provide an 'estimate' for the change score in $P_1$. Suppose that $x_{hj}$ is the change score and $sd_{hj}$ the standard deviation provided by the expert $h = 1,2, \dots, H$ for treatment $j \in T_c$. Let also $\gamma_h$ denote the overall confidence of each expert to provided opinions. To combine the estimates provided by different experts, we can then use the following meta-analysis model:

$$x_{hj} \sim \mathrm{N}\left(c_{hj}, \frac{sd_{hj}^2}{\gamma_h}\right) \tag{7}$$

$$d_{hj} \sim \mathrm{N}(\xi_j^{P_1}, \sigma^2) \tag{8}$$

with $d_{hj} = \frac{c_{hj}}{\mathrm{med}(sd_i^{pooled})}, \forall i, h, j$ the standardized values of $c_{hj}$. Given that the pooled standard deviation cannot be elicited by experts, the median across the studies in the data is used. Note that standardization is not necessary when all studies in the data use the same scale. Then, the pooled SMDs can be obtained as $u_{1j}^{P_1} = \xi_j^{P_1} - \xi_1^{P_1}$ and the priors for the parameter $\beta_{1j}$ is constructed as in Equation (6).

*Prior distributions for the scale parameter*

Dividing the variances in Equation (4) by $w_i$ can be seen as a special case of the power prior method where a specific power is chosen for the likelihood of each study[9,13,14]. Values of $w_i$ close to 0 reflect a serious downweight of the evidence coming from the external subgroup $P_2$ while values close to 1 imply a mild downweight. Typical choices of prior distributions for the scale parameter can be the beta or the uniform distribution. The parameters of these distributions should be chosen to reflect the prior beliefs regarding the specific characteristic according to which the downweight takes place. For example, a $\mathrm{Beta}(3,3)$ can be used for cases where a moderate downweight needs to apply for the external information. This distribution places its mass around 0.5 with 95% range equal to [0.15, 0.84] thus reflecting more uncertainty about the magnitude of our downweighting. Left-skewed beta priors (e.g. $\mathrm{Beta}(1,6)$) can be used for weight values close to 0 while on the other hand right-skewed beta priors (e.g. $\mathrm{Beta}(6,1)$) can be used for weight values close to 1. The previous downweighting schemes can be achieved with other types of prior distributions. For example, a $\mathrm{Unif}(0.4,0.6)$, a $\mathrm{Unif}(0.1,0.3)$ or a $\mathrm{Unif}(0.8,1)$ can be used for moderate, serious or mild downweight, respectively. Finally, it is generally not recommended to use priors for the scale parameter which can allow for values larger than 1. This could further increase the impact of external data and enable their dominance in the final NMA estimates.

# 4. Application

## 4.1 Implementation

We consider throughout CA being the target population subgroup $P_1$ and GP the external subgroup $P_2$. We applied in total 9 models 6 of which refer to the combinations of the different prior distributions for the location ($\beta_{1j}$) and scale parameters ($w_i$).

1. Naïve synthesis model for CA and GP with non-informative priors
2. NMA for CA with informative priors from GP
    a. data-based prior distributions for the location parameters with
        i. no downweight for any GP study ($w_i = 1$) using only treatments in $T_c$ (No DW)
        ii. moderate downweight (i.e. $w_i \sim \text{Beta}(3,3)$) to high risk of bias (RoB) GP studies using only treatments in $T_c$ (RoB DW). The risk of bias was assessed according to the Cochrane's RoB tool[15] and in total 16 studies were rated at high RoB for GP.
        iii. no downweight for any GP study evaluating treatments in $T_c$ and moderate downweight (i.e. $w_i \sim \text{Beta}(3,3)$) for those evaluating at least one treatment in $T_a - T_c$ (NCT DW). This means that here we use the full network of GP with the 34 treatments and we downweight all studies that contain at least one treatment that has never been evaluated in CA (107 studies).
    b. prior distributions based on expert opinion for the location parameter combined the three above (i-iii) possibilities for the scale parameters
3. NMA for CA with non-informative priors (i.e. 'standard' NMA)
4. Pairwise meta-analysis for CA with non-informative priors

All the analyses were conducted using the `rjags`[16] package through the `R` statistical software (`R` version 4.0.3, 2020-10-10)[17]. For all models, we ran two chains in parallel, performed 50,000 iterations and discarded the first 10,000 samples of each chain. We checked the chains convergence using the Gelman-Rubin criterion[18] with a value below 1.1 to indicate failure of convergence. We also visually checked the Markov chains using trace plots to inspect the sampling behavior and assess mixing across chains and convergence.

## 4.2 Elicitation of expert's opinion

We prepared a questionnaire and circulated it to psychiatrists with experience in treating schizophrenia for CA and GP (available in Appendix 1). We asked each expert to provide an estimate of the 'expected' mean score reduction in the PANSS scale from baseline to endpoint for each drug for CA given the pooled mean score reduction obtained from the data for GP. We further asked them to provide a measure of uncertainty around the given expected mean scores a) in the form of standard deviation and b) in a 10-point scale of their confidence. We obtained in total 22 responses which we averaged for each drug using the meta-analysis model given in Equations (7)-(8). We used a non-informative $N(0{,}10000)$ and a weakly-informative HN(1) prior for the means $\xi_j^{P_1}, j \in T_c$ and the parameter $\sigma$ respectively.

## 4.3 Results

The estimates for the basic comparisons of the CA network are depicted in **Figure 2** and for all relative comparisons in Appendix 2, Tables 9-16. As expected, for most comparisons the use of informative priors leads to a substantial improvement in the precision of the relative effects. The naïve synthesis model provides the most precise results but relies on a strong assumption that is likely to be implausible. Interestingly, CA results from the standard NMA model (with non-informative priors) tend to be less precise than the respective direct estimates. This occurs often in sparse networks since indirect comparisons as well as heterogeneity are estimated with large uncertainty. Overall, standard NMA most often does not give any insight for comparisons without direct evidence as it yields very large credible intervals. For some of these comparisons (e.g. Haloperidol vs Placebo) the NMA models with informative priors result in more conclusive relative effect estimates. In terms of point estimates, results appear generally robust across the different models with only few exceptions, such as the extreme cases of Fluphenazine and Trifluoperazine. Those two are very old drugs for which the evidence is outdated and sparse in both GP and CA networks. The latter implies that the informative prior can dominate in the analysis and yield to estimate closer to the GP. Overall, point estimates appeared to be more robust for the basic comparisons for which direct evidence is available in the network of CA. Finally, no convergence issues were identified across the chains. The corresponding trace plots are available in Appendix 2.

Only small differences can be observed among the six models with informative priors suggesting that the approach of informing the location and the scale parameters does not affect materially the final predictions; this is possibly due to the huge amount of data in the GP network. For a few comparisons the data-based approach for $\boldsymbol{\beta}$ provides to some degree different relative effects than those obtained from the expert opinion approach implying that the experience from clinical practice does not fully agree with the available data. Those comparisons might need to be prioritized for the design of new trials. Of course, when the full GP network was used and only studies comparing treatments in $T_a - T_c$ were downweighted some extra precision was gained. Such an approach might be more useful when the external network is not as dense as in the present dataset.

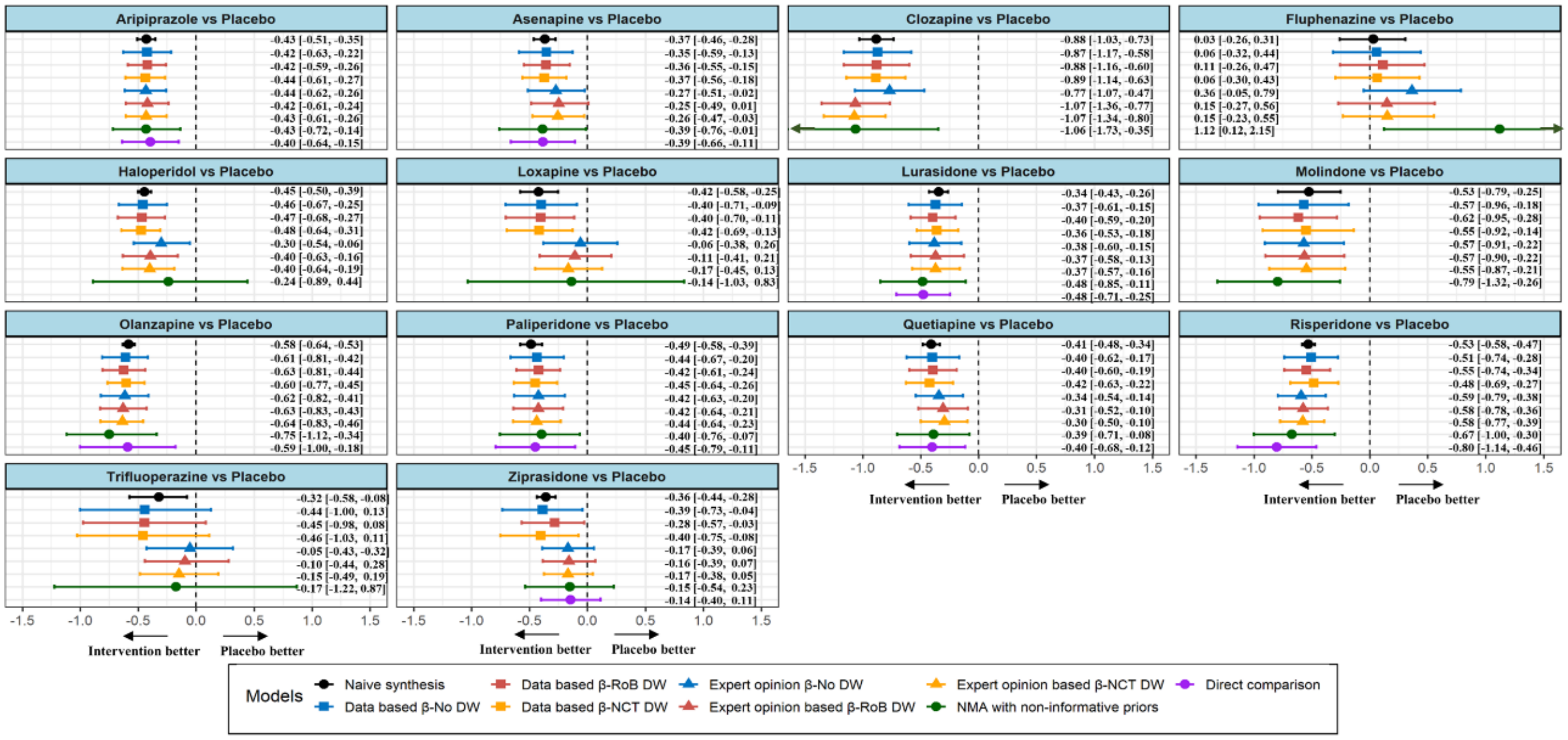

**Figure 2:** Results from the analysis of the network defined for the population of CA. The outcome of interest is the reduction of the overall schizophrenia symptoms and the effect measure is the standardized mean difference.

Assessment of inconsistency was performed using the node-splitting method[5]. Inference regarding the presence of inconsistency was based on Bayesian p-values and on the visual inspection of the posterior densities of direct and indirect evidence, Bayesian p-values were calculated as $2 \times \min\{P, 1 - P\}$ where $P$ is the probability that the difference between direct and indirect comparisons is positive. Additionally, the different antipsychotics and placebo were ranked across all models using the surface under the cumulative ranking curve (SUCRA)[19] method. Overall, the respective posterior densities of the direct and indirect estimates appeared to be quite similar across the different comparisons and models and no p-value indicated lack of consistency (Appendix 2, Figures 1-5, Tables 1-8). In terms of ranking, the results appear to be generally robust across models including the naïve synthesis and the standard NMA model. Clozapine was always ranked as the most effective antipsychotic. Finally, Olanzapine, Risperidone and Molindone were ranked in second, third and fourth position respectively in most of the fitted models (Appendix 2, Figure 6).

# 5. Discussion

In this paper, we present a Bayesian framework for NMA of sparse networks aiming to improve the performance of the relative effect estimates in terms of precision and reliability. To this end, we borrow information using external data from a dense network that targets a different subgroup of the population. We model the differences between the two populations with a two-stage approach. At the first stage, we analyze only the dense network and we extrapolate its results to the sparse network though the incorporation of a location parameter in the distribution of the summary relative effects. Then, we use the extrapolated results to construct informative prior distributions for the target population subgroup. Similar approaches have been used previously at the level of primary studies[7,20–22]. We further introduce a scale parameter that inflates the variance of the studies in the dense network to reflect the potential uncertainty regarding the assumed relationship between the two population subgroups[23,24].

We proposed two different approaches for informing the location parameters; one based on the data and one based on expert opinion. The former is easier to implement but the data from the sparse network are used in both stages and this might introduce some correlation that is ignored. In addition, this approach requires that a certain number of direct comparisons is available in both networks, otherwise the use of non-informative priors would be necessary for several $\beta$s at the first stage. The prior elicitation approach is general as it does not rely on the amount of direct evidence. However, results from such an approach are always subjective to some degree. Here, we requested the necessary information from the psychiatrists though a questionnaire and some of them raised concerns that it was difficult for them to make a suggestion. Other ways to elicit information from experts exist. For example, Turner et al. used a three stage elicitation approach where the experts were interviewed through 1:1 meetings or by telephone[25].

In our motivating example, the two approaches for constructing priors for the location parameters were generally in agreement. For the scale parameters, a moderate or no downweight was chosen mostly for the purpose of demonstrating our approach. Compared to the standard NMA with non-informative priors, our two-stage approach provided substantially more precise estimates. The lack of robustness of the standard NMA model for this dataset is also deduced from the fact that almost always it yielded less precise results than pairwise meta-analysis. As expected, the most precise results were obtained from the naïve synthesis model. Nevertheless, this model does not account for any differences across the two population subgroups.

In the presence of few studies, estimation of the heterogeneity is challenging[8]. In our analysis we used a weakly-informative half normal prior distribution for the heterogeneity parameter[26]. Using informative priors for heterogeneity could possibly further increase the precision of our approach. Usually, these are based on empirical distributions that depend on the type of outcome and treatment comparisons[27–29]. Alternatively, more sophisticated options that allow the incorporation of external data in the estimation of the heterogeneity parameter may be applied[8].

Overall, the proposed framework offers a reliable approach for assisting the analysis of sparse networks. Of course, its application requires close collaboration between statisticians and experienced clinicians to ensure that sharing of information between different population subgroups is clinically meaningful. Finally, conducting NMA in cases of sparse networks will always be a challenging procedure and the best approach will likely be context-dependent.

**Acknowledgements:** We are thankful to the following psychiatrists for providing expert opinion: Arango Celso, Carpenter Will, Crespo Facorro Benedicto, Davidson Michael, Dazzan Paola, Gerhard Gruender, Hasan Alkomiet, Heres Stephan, Hui Wu, Jauhar Sameer, Juckel Georg, McGorry Pat, Murray Robin, Musil Richard, Peter Natalie, Samara Myrto, Schneider-Thoma Johannes, Siafis Spyridon, Stip Emmanuel, Takeuchi Hiroyoshi, Weiser Mark, Zhu Yikang. RT was supported by the UK Medical Research Council (Programme MC_UU_00004/06).

**Conflicts of interest:** The authors have declared no conflicts of interest

Questionnaire for methodological project on the analysis of psychiatric symptom reduction by antipsychotics in children and adolescents

**Short introduction to the project (please read this before starting answering the questions)**

In this project we aim to estimate the effects of different antipsychotics for children and adolescents (Krause et al. 2018[1]). However, the available evidence for this population is very limited. To increase the confidence in our results, we plan to borrow information from available evidence on the same outcome for "general" patients (usually chronic adults with an exacerbation of positive symptoms, for example as those meta-analyzed by Huhn et al. 2019[2]) for whom more evidence is available. Thus, we need to estimate the 'difference' between the two populations. To do this properly, we would need your help by answering a questionnaire that we have prepared below.

More details about our methodology can be found in the following section "**Methodological details related to the project**".
Otherwise, please go directly to the questionnaire in the section "

**Questions to be answered by experts**" in page 4.

## Methodological details related to the project

Sparse data are a common issue in comparative effectiveness research even at the meta-analysis level. We use as an exemplar sparse network, the network formed by 19 RCTs comparing 14 antipsychotics and placebo for the reduction of overall psychiatric symptoms (PANSS total score) in children and adolescents. This network is depicted in **Figure 1**. The 19 RCTs provide 21 out of the 105 (20%) theoretically possible direct comparisons (lines in the graph). Performing a 'standard' network meta-analysis (NMA) in this network, which combines direct (example: all studies comparing aripiprazole and paliperidone with each other) and indirect evidence (example: estimating the difference between aripiprazole and paliperidone indirectly from aripiprazole versus placebo and paliperidone versus placebo), is possibly of limited validity. Specifically, the resulting relative effects are equally or even less precise for some comparisons than those obtained from pairwise meta-analysis and the model assumptions cannot be evaluated. However, the pace of new research production for this network is very slow and it is of great importance to use any available piece of information to identify the most beneficial drugs for children and adolescents.

The main idea behind the following questionnaire is that we could use external evidence from a dense (i.e. informative) network after some form of 'adaptation' of the results to the population of interest and increase the amount of information for the sparse network for children and adolescents. In other words, we could use this adapted external evidence as 'prior' information in a Bayesian NMA model for children and adolescents. In this way, we aim to obtain more precise and more robust (reliable) results for the comparative efficacy of the drugs for children and adolescents.

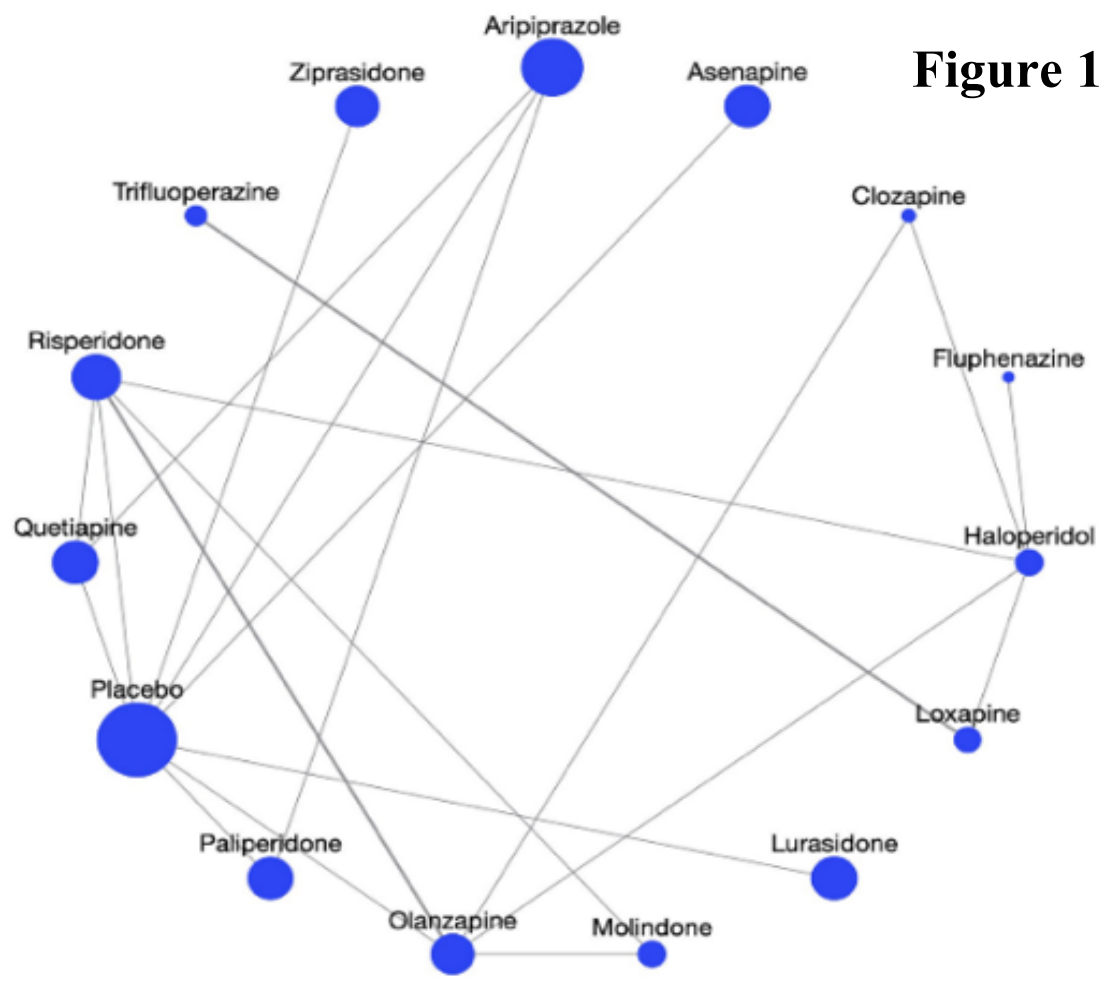

Figure 1

Figure 2

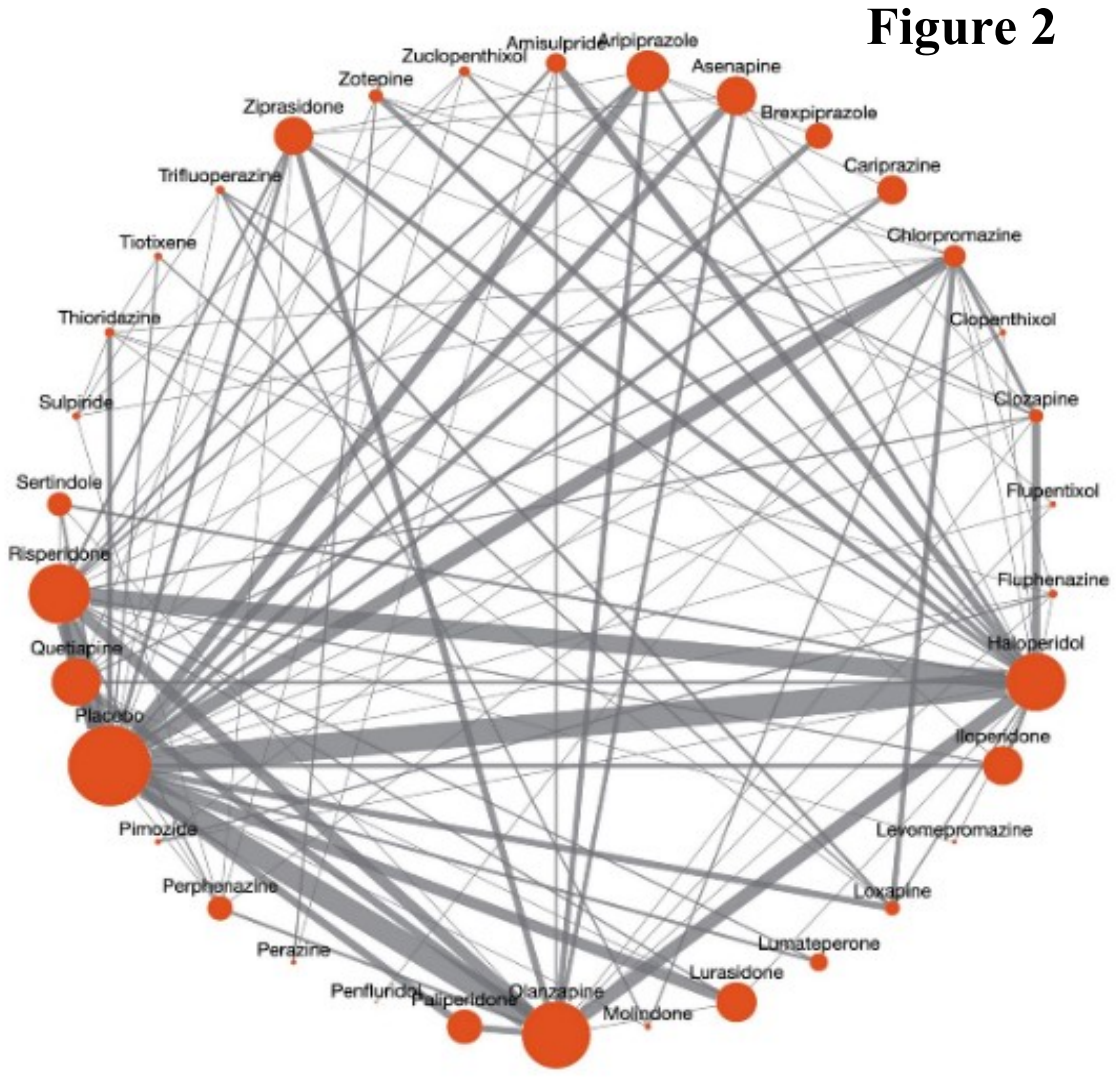

The external informative network we consider here is a network evaluating the drugs for the same outcome (overall symptoms) in "general" adult patients with schizophrenia (**Figure 2**) as they were, for example, meta-analyzed in Huhn et al. 2019[2]. These studies, albeit they do not always use exactly the same definitions, can be characterised as usually chronic patients with an acute exacerbation of positive symptoms. This network consists of 255 studies and 116 direct comparisons. Note that we will only use the evidence from this network for the drugs already existing in the children and adolescent network; hence we will not draw any inference for drugs that have never been used for the population of interest. Of course, some differences among the two populations are inevitable and, therefore, we cannot use the results from the general population without incorporating these differeces. Since we only have aggregate (i.e. summary) data from the studies, we will rely on the average difference between the outcome in the two populations for each drug.

From a statistical standpoint, the average difference in the outcome between the two populations is represented by the difference of the mean of the respective distributions. **Figure 3** shows a theoretical example about the difference of the means of the two distributions for the comparison haloperidol vs placebo. Such differences for all comparisons can be obtained either using the data (which we will do, as well) or using expert opinion (this is what we need your help for). Then, these differences will be modelled appropriately to obtain NMA results from the general patient data applicable to children and adolescents which, in turn, will be used as prior information in the NMA for children and adolescents.

**Figure 3**

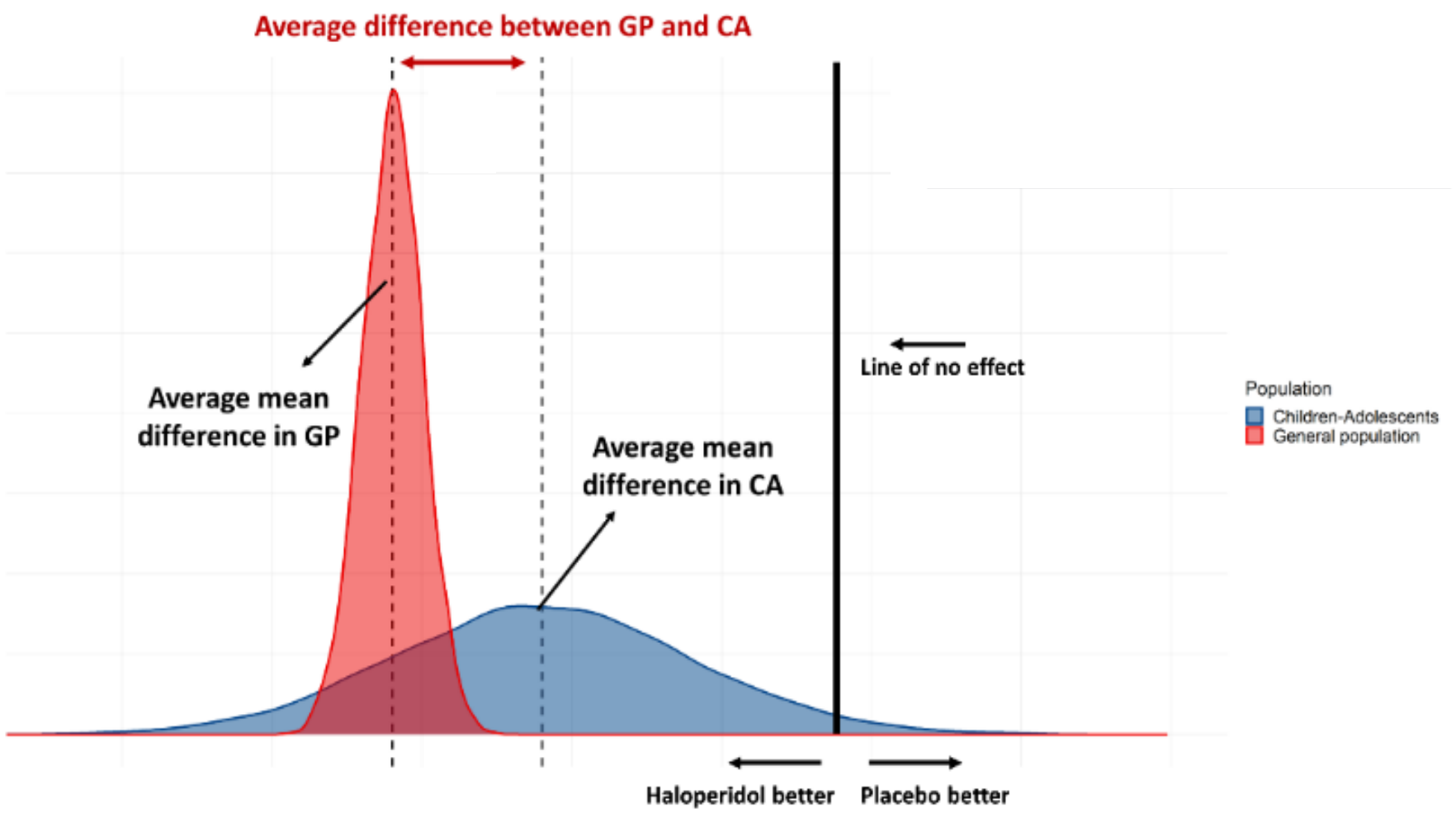

## Questions to be answered by experts

**Q1:** Column 2 of the next table shows a theoretical reduction (change score) in PANSS total score for overall symptoms for the antipsychotics of Figure 2 in "general patients" (chronic adults with acute exacerbation of positive symptoms).

***Please fill into column 3 how much you expect that the corresponding reductions would be for children and adolescents. Please, also provide a standard deviation for the value you provide in column 4.***

You may skip the antipsychotics which you are not familiar with.

***Please also tick the box in column 5 regarding how you think the reduction of the PANSS total score in children and adolescents <u>qualitatively</u> differs from that in general patients.***

**<u>The first row in red presents a fictional example:</u>** If the PANSS total score reduction from baseline to endpoint for "general patients" for a given fictional antipsychotic X is -26 then a possible answer for children and adolescents may be:

- Expected PANSS total score reduction from baseline to endpoint in Children-Adolescents=-29 meaning three points more improvement
- Standard Deviation for the expected PANSS reduction in Children-Adolescents =±10
- Expected treatment response in qualitative terms in Children-Adolescents compared to "general patients" = A bit better

| 1. Drug | 2. PANSS total score reduction from baseline to endpoint for "general adult patients" (chronic adults with acute exacerbation of positive symptoms) | 3. Expected PANSS total score reduction from baseline to endpoint in Children-Adolescents | 4. Standard deviation for the expected PANSS reduction in Children-Adolescents | 5. Expected treatment response in qualitative terms in Children-Adolescents compared to "general adult patients" |
|---|---|---|---|---|
| Fictional antipsychotic X | -26 | -29 | ±10 | ☐ Much better ☒ A bit better ☐ Same ☐ A bit worse ☐ Much worse |
| Clozapine | -27.8 | | | ☐ Much better ☐ A bit better ☐ Same ☐ A bit worse ☐ Much worse |
| Olanzapine | -21.2 | | | ☐ Much better ☐ A bit better ☐ Same ☐ A bit worse ☐ Much worse |
| **Risperidone** | -21 | | | ☐ Much better ☐ A bit better ☐ Same ☐ A bit worse ☐ Much worse |
| **Paliperidone** | -19.8 | | | ☐ Much better ☐ A bit better ☐ Same ☐ A bit worse ☐ Much worse |
| **Haloperidol** | -19.4 | | | ☐ Much better ☐ A bit better ☐ Same ☐ A bit worse ☐ Much worse |
| **Loxapine** | -19 | | | ☐ Much better ☐ A bit better ☐ Same ☐ A bit worse ☐ Much worse |
| **Quetiapine** | -18.4 | | | ☐ Much better ☐ A bit better ☐ Same ☐ A bit worse ☐ Much worse |
| **Molindone** | -18.4 | | | ☐ Much better ☐ A bit better ☐ Same ☐ A bit worse ☐ Much worse |
| **Aripiprazole** | -18.2 | | | ☐ Much better ☐ A bit better ☐ Same ☐ A bit worse ☐ Much worse |
| **Ziprasidone** | -18.2 | | | ☐ Much better ☐ A bit better ☐ Same ☐ A bit worse ☐ Much worse |
| **Asenapine** | -17.8 | | | ☐ Much better ☐ A bit better ☐ Same ☐ A bit worse ☐ Much worse |
| **Lurasidone** | -17.2 | | | ☐ Much better ☐ A bit better ☐ Same ☐ A bit worse ☐ Much worse |
| **Fluphenazine** | -14.8 | | | ☐ Much better ☐ A bit better ☐ Same ☐ A bit worse ☐ Much worse |
| **Trifluoperazine** | -14.8 | | | ☐ **Much better** ☐ **A bit better** ☐ **Same** ☐ **A bit worse** ☐ **Much worse** |
| **Placebo** | -10 | | | ☐ **Much better** ☐ **A bit better** ☐ **Same** ☐ **A bit worse** ☐ **Much worse** |

**Q2:** In a scale from 1 to 10 how confident you are in the numbers that you gave in **Q1**? (1=not confident, 10=very confident)

| **1** | **2** | **3** | **4** | **5** | **6** | **7** | **8** | **9** | **10** |
|---|---|---|---|---|---|---|---|---|---|
| ☐ | ☐ | ☐ | ☐ | ☐ | ☐ | ☐ | ☐ | ☐ | ☐ |

**Q3:** What are the most probable population related factors that may differentiate the effectiveness of the antipsychotics between general patients and children and adolescents?

**Q4:** Consider that a new drug with very positive results for the general population has come out (very efficacious without important side effects) but not tested in children and adolescents. Would you consider it for treating children/adolescents with schizophrenia? If yes, in which situations (please fill-in the box below)? **Please disregard any legal concerns with off-label use.**

Certainly yes ☐

Possibly yes ☐

Maybe ☐

Possibly no ☐

Certainly no ☐

**If not, why would you <u>not</u> consider giving the potential new drug to children and adolescents when evidence is available only for the general population?**

**Q5:** Your main work is

Research ☐

Clinical practice ☐

Both ☐

**Q6:** How many years have you been working in the field of schizophrenia?

| |
|---|
| |

**Q7:** You have experience with

Child psychiatry ☐

Adult psychiatry ☐

Both ☐

**Q8:** What kind of experience do you have with studies about antipsychotics?

Involved in randomized clinical trials ☐

Involved in meta-analyses of clinical trials ☐

Involved in observational studies ☐

Any other study related experience ☐

# Appendix 2

## Consistency checks

**Appendix Table 1:** Consistency checks for the naive synthesis model.

| Comparison | Direct [95% CI] | Indirect [95% CI] | Difference [95% CI] | P(diff>0) | p-value |
|---|---|---|---|---|---|
| Aripiprazole vs Placebo | -0.39 [-0.50, -0.27] | -0.46 [-0.56, -0.36] | 0.07 [-0.07, 0.22] | 0.84 | 0.31 |
| Olanzapine vs Placebo | -0.50 [-0.58, -0.42] | -0.59 [-0.66, -0.51] | 0.09 [-0.02, 0.20] | 0.94 | 0.12 |
| Paliperidone vs Placebo | -0.52 [-0.65, -0.39] | -0.53 [-0.66, -0.40] | 0.01 [-0.19, 0.21] | 0.51 | 0.98 |
| Quetiapine vs Placebo | -0.30 [-0.41, -0.19] | -0.41 [-0.52, -0.30] | 0.11 [-0.05, 0.26] | 0.90 | 0.20 |
| Risperidone vs Placebo | -0.44 [-0.52, -0.35] | -0.50 [-0.59, -0.41] | 0.06 [-0.06, 0.19] | 0.85 | 0.29 |

**Appendix Table 2:** Consistency checks for the NMA model with informative priors obtained from GP using a data based approach for $\beta$ and no downweight.

| Comparison | Direct [95% CI] | Indirect [95% CI] | Difference [95% CI] | P(diff>0) | p-value |
|---|---|---|---|---|---|
| Aripiprazole vs Placebo | -0.40 [-0.60, -0.19] | -0.50 [-0.82, -0.18] | 0.10 [-0.27, 0.47] | 0.71 | 0.57 |
| Olanzapine vs Placebo | -0.58 [-0.80, -0.37] | -0.73 [-1.05, -0.39] | 0.15 [-0.27, 0.55] | 0.77 | 0.46 |
| Paliperidone vs Placebo | -0.45 [-0.71, -0.22] | -0.34 [-0.76, 0.09] | -0.11 [-0.62, 0.36] | 0.32 | 0.64 |
| Quetiapine vs Placebo | -0.41 [-0.63, -0.17] | -0.33 [-0.74, 0.11] | -0.07 [-0.55, 0.41] | 0.38 | 0.76 |
| Risperidone vs Placebo | -0.57 [-0.80, -0.32] | -0.49 [-0.85, -0.14] | -0.08 [-0.50, 0.36] | 0.35 | 0.71 |

**Appendix Table 3:** Consistency checks for the NMA model with informative priors obtained from GP using a data based approach for $\beta$ and moderate downweight to all GP studies with high RoB.

| Comparison | Direct [95% CI] | Indirect [95% CI] | Difference [95% CI] | P(diff>0) | p-value |
|---|---|---|---|---|---|
| Aripiprazole vs Placebo | -0.39 [-0.60, -0.20] | -0.50 [-0.80, -0.19] | 0.10 [-0.27, 0.46] | 0.71 | 0.57 |
| Olanzapine vs Placebo | -0.58 [-0.79, -0.38] | -0.73 [-1.09, -0.38] | 0.15 [-0.26, 0.55] | 0.77 | 0.46 |
| Paliperidone vs Placebo | -0.45 [-0.67, -0.23] | -0.35 [-0.74, 0.05] | -0.10 [-0.56, 0.35] | 0.32 | 0.64 |
| Quetiapine vs Placebo | -0.42 [-0.64, -0.18] | -0.33 [-0.71, 0.06] | -0.09 [-0.54, 0.36] | 0.35 | 0.71 |
| Risperidone vs Placebo | -0.57 [-0.80, -0.31] | -0.50 [-0.86, -0.14] | -0.06 [-0.51, 0.39] | 0.38 | 0.76 |

**Appendix Table 4:** Consistency checks for the NMA model with informative priors obtained from GP using a data based approach for $\beta$ and moderate downweight to all GP studies with interventions in $T_a - T_c$.

| Comparison | Direct [95% CI] | Indirect [95% CI] | Difference [95% CI] | P(diff>0) | p-value |
|---|---|---|---|---|---|
| Aripiprazole vs Placebo | -0.41 [-0.57, -0.23] | -0.51 [-0.81, -0.20] | 0.10 [-0.26, 0.44] | 0.73 | 0.54 |
| Olanzapine vs Placebo | -0.59 [-0.76, -0.41] | -0.73 [-1.07, -0.39] | 0.14 [-0.23, 0.53] | 0.76 | 0.48 |
| Paliperidone vs Placebo | -0.46 [-0.66, -0.26] | -0.36 [-0.74, 0.04] | -0.10 [-0.54, 0.32] | 0.31 | 0.62 |
| Quetiapine vs Placebo | -0.43 [-0.64, -0.22] | -0.35 [-0.75, 0.05] | -0.08 [-0.53, 0.36] | 0.36 | 0.71 |
| Risperidone vs Placebo | -0.54 [-0.74, -0.32] | -0.49 [-0.84, -0.12] | -0.05 [-0.46, 0.36] | 0.41 | 0.81 |

**Appendix Table 5:** Consistency checks for the NMA model with informative priors obtained from GP using expert's opinion for $\beta$ and no downweight.

| Comparison | Direct [95% CI] | Indirect [95% CI] | Difference [95% CI] | P(diff>0) | p-value |
|---|---|---|---|---|---|
| Aripiprazole vs Placebo | -0.41 [-0.65, -0.19] | -0.47 [-0.80, -0.14] | 0.06 [-0.35, 0.47] | 0.62 | 0.77 |
| Olanzapine vs Placebo | -0.58 [-0.83, -0.33] | -0.69 [-1.04, -0.33] | 0.11 [-0.31, 0.54] | 0.71 | 0.58 |
| Paliperidone vs Placebo | -0.47 [-0.74, -0.21] | -0.34 [-0.75, 0.08] | -0.13 [-0.62, 0.35] | 0.30 | 0.59 |
| Quetiapine vs Placebo | -0.33 [-0.57, -0.07] | -0.37 [-0.75, 0.06] | 0.04 [-0.45, 0.52] | 0.55 | 0.90 |
| Risperidone vs Placebo | -0.66 [-0.88, -0.42] | -0.42 [-0.77, -0.06] | -0.24 [-0.68, 0.19] | 0.15 | 0.29 |

**Appendix Table 6:** Consistency checks for the NMA model with informative priors obtained from GP using expert's opinion for $\beta$ and moderate downweight to all GP studies with high RoB.

| Comparison | Direct [95% CI] | Indirect [95% CI] | Difference [95% CI] | P(diff>0) | p-value |
|---|---|---|---|---|---|
| Aripiprazole vs Placebo | -0.41 [-0.64, -0.17] | -0.45 [-0.77, -0.10] | 0.05 [-0.37, 0.44] | 0.59 | 0.81 |
| Olanzapine vs Placebo | -0.57 [-0.81, -0.32] | -0.76 [-1.11, -0.41] | 0.20 [-0.22, 0.63] | 0.81 | 0.37 |
| Paliperidone vs Placebo | -0.47 [-0.72, -0.20] | -0.32 [-0.73, 0.07] | -0.14 [-0.62, 0.33] | 0.27 | 0.55 |
| Quetiapine vs Placebo | -0.28 [-0.52, -0.04] | -0.35 [-0.79, 0.07] | 0.08 [-0.43, 0.58] | 0.61 | 0.78 |
| Risperidone vs Placebo | -0.63 [-0.85, -0.39] | -0.45 [-0.80, -0.07] | -0.18 [-0.60, 0.24] | 0.20 | 0.40 |

**Appendix Table 7:** Consistency checks for the NMA model with informative priors obtained from GP using expert's opinion for $\beta$ and moderate downweight to all GP studies with interventions in $T_a - T_c$.

| Comparison | Direct [95% CI] | Indirect [95% CI] | Difference [95% CI] | P(diff>0) | p-value |
|---|---|---|---|---|---|
| Aripiprazole vs Placebo | -0.42 [-0.62, -0.21] | -0.46 [-0.77, -0.13] | 0.04 [-0.34, 0.41] | 0.59 | 0.82 |
| Olanzapine vs Placebo | -0.58 [-0.82, -0.36] | -0.74 [-1.09, -0.40] | 0.16 [-0.25, 0.58] | 0.78 | 0.44 |
| Paliperidone vs Placebo | -0.48 [-0.73, -0.24] | -0.33 [-0.71, 0.08] | -0.16 [-0.65, 0.29] | 0.25 | 0.49 |
| Quetiapine vs Placebo | -0.27 [-0.48, -0.04] | -0.36 [-0.78, 0.07] | 0.10 [-0.38, 0.56] | 0.65 | 0.69 |
| Risperidone vs Placebo | -0.62 [-0.83, -0.39] | -0.44 [-0.81, -0.08] | -0.18 [-0.60, 0.25] | 0.20 | 0.40 |

**Appendix Table 8:** Consistency checks for the NMA model with non-informative priors.

| Comparison | Direct [95% CI] | Indirect [95% CI] | Difference [95% CI] | P(diff>0) | p-value |
|---|---|---|---|---|---|
| Aripiprazole vs Placebo | -0.40 [-0.86, 0.07] | -0.49 [-0.95, 0.02] | 0.09 [-0.62, 0.76] | 0.63 | 0.74 |
| Olanzapine vs Placebo | -0.58 [-1.10, -0.05] | -0.96 [-1.51, -0.39] | 0.38 [-0.43, 1.19] | 0.85 | 0.30 |
| Paliperidone vs Placebo | -0.44 [-0.95, 0.12] | -0.35 [-1.02, 0.27] | -0.09 [-0.85, 0.71] | 0.40 | 0.79 |
| Quetiapine vs Placebo | -0.40 [-0.88, 0.07] | -0.37 [-0.92, 0.19] | -0.02 [-0.75, 0.69] | 0.47 | 0.95 |
| Risperidone vs Placebo | -0.80 [-1.21, -0.38] | -0.49 [-1.05, 0.03] | -0.32 [-0.99, 0.34] | 0.15 | 0.31 |

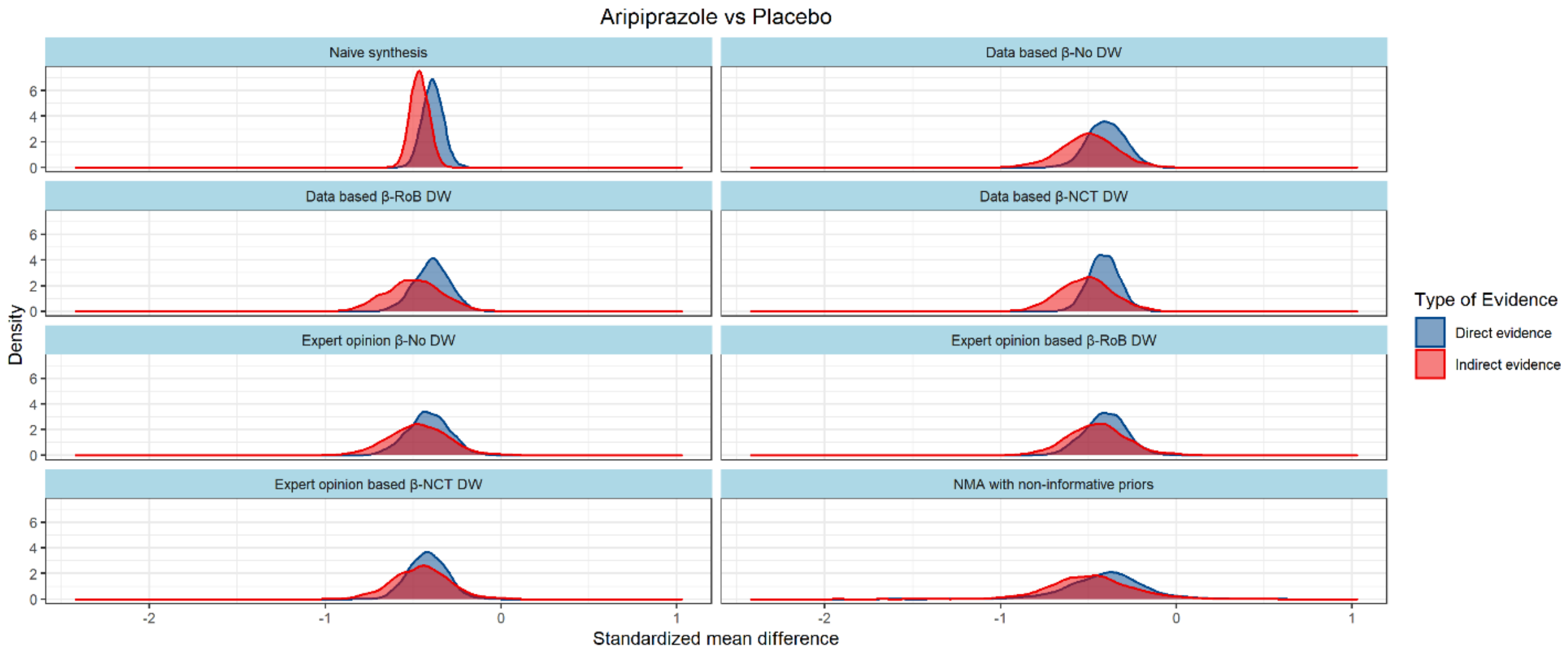

**Appendix Figure 1:** Posterior densities for direct and indirect evidence in terms of comparison Aripiprazole versus Placebo across all the different models.

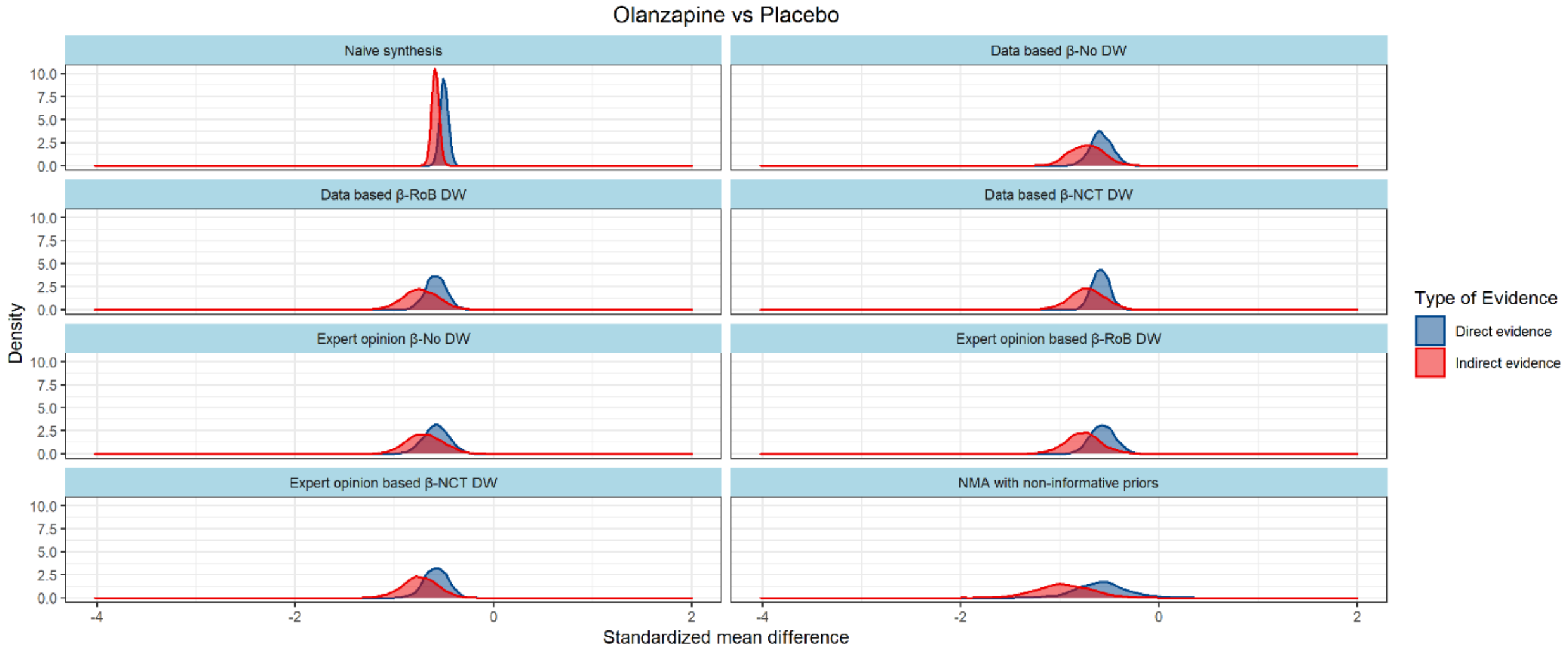

**Appendix Figure 2:** Posterior densities for direct and indirect evidence in terms of comparison Olanzapine versus Placebo across all the different models.

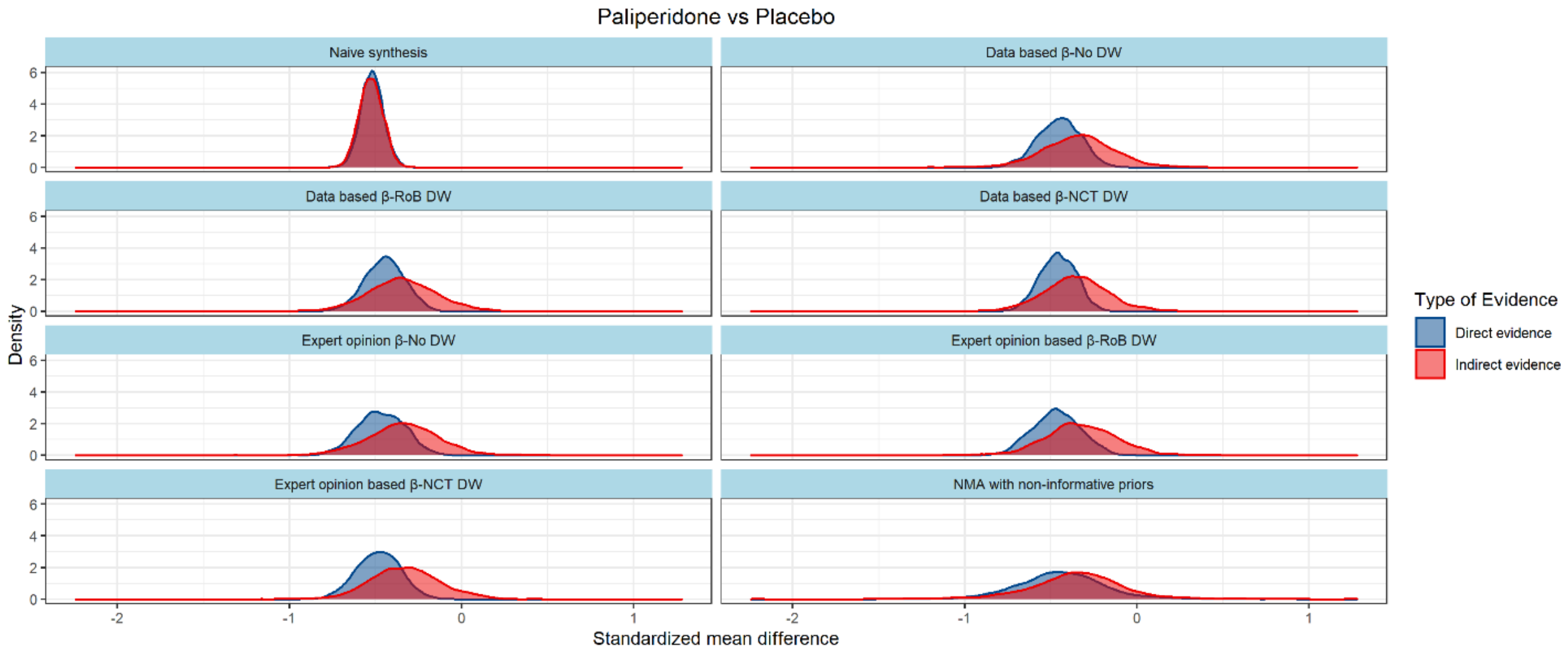

**Appendix Figure 3:** Posterior densities for direct and indirect evidence in terms of comparison Paliperidone versus Placebo across all the different models.

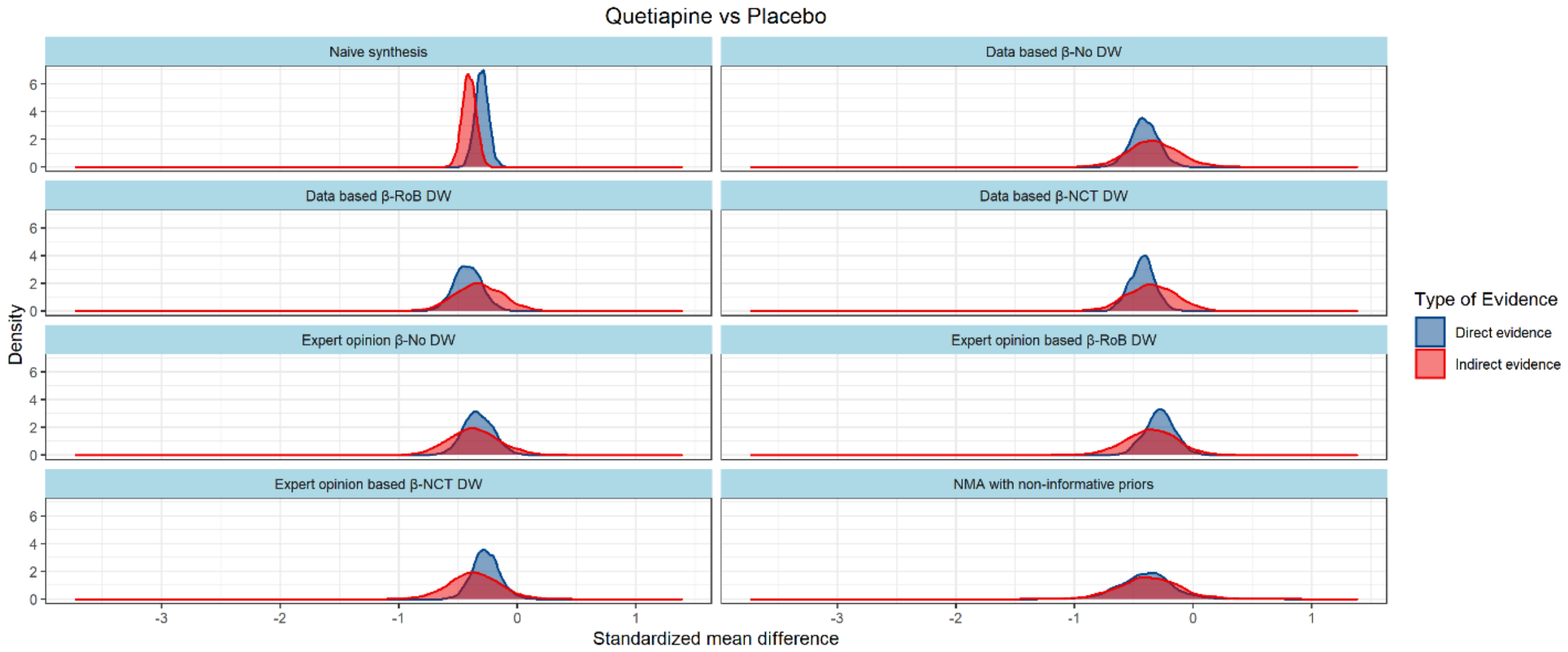

**Appendix Figure 4:** Posterior densities for direct and indirect evidence in terms of comparison Quetiapine versus Placebo across all the different models.

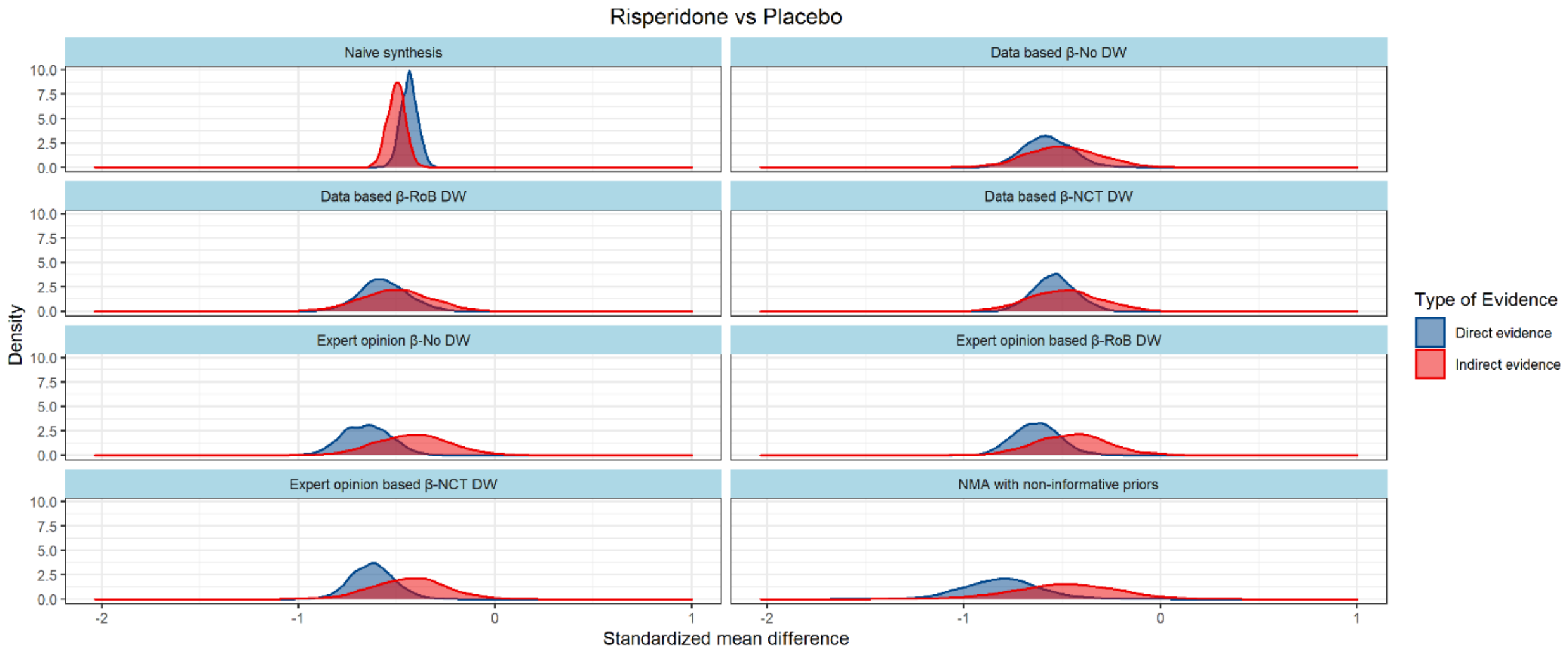

**Appendix Figure 5:** Posterior densities for direct and indirect evidence in terms of comparison Risperidone versus Placebo across all the different models.

# Ranking between treatments across all models

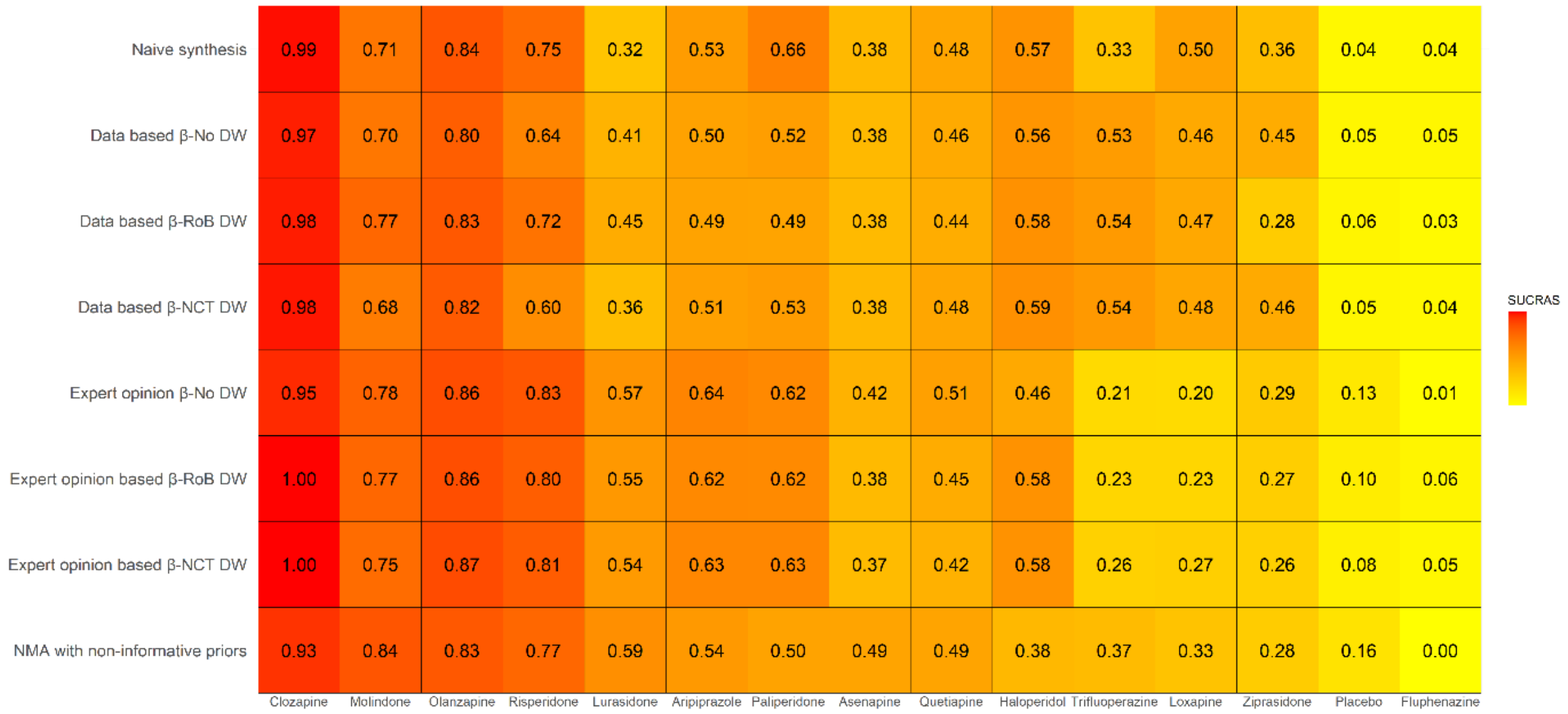

**Appendix Figure 6:** Ranking between all the drugs (14 antipsychotics and Placebo) as obtained from each one of the different models.

## Traceplots

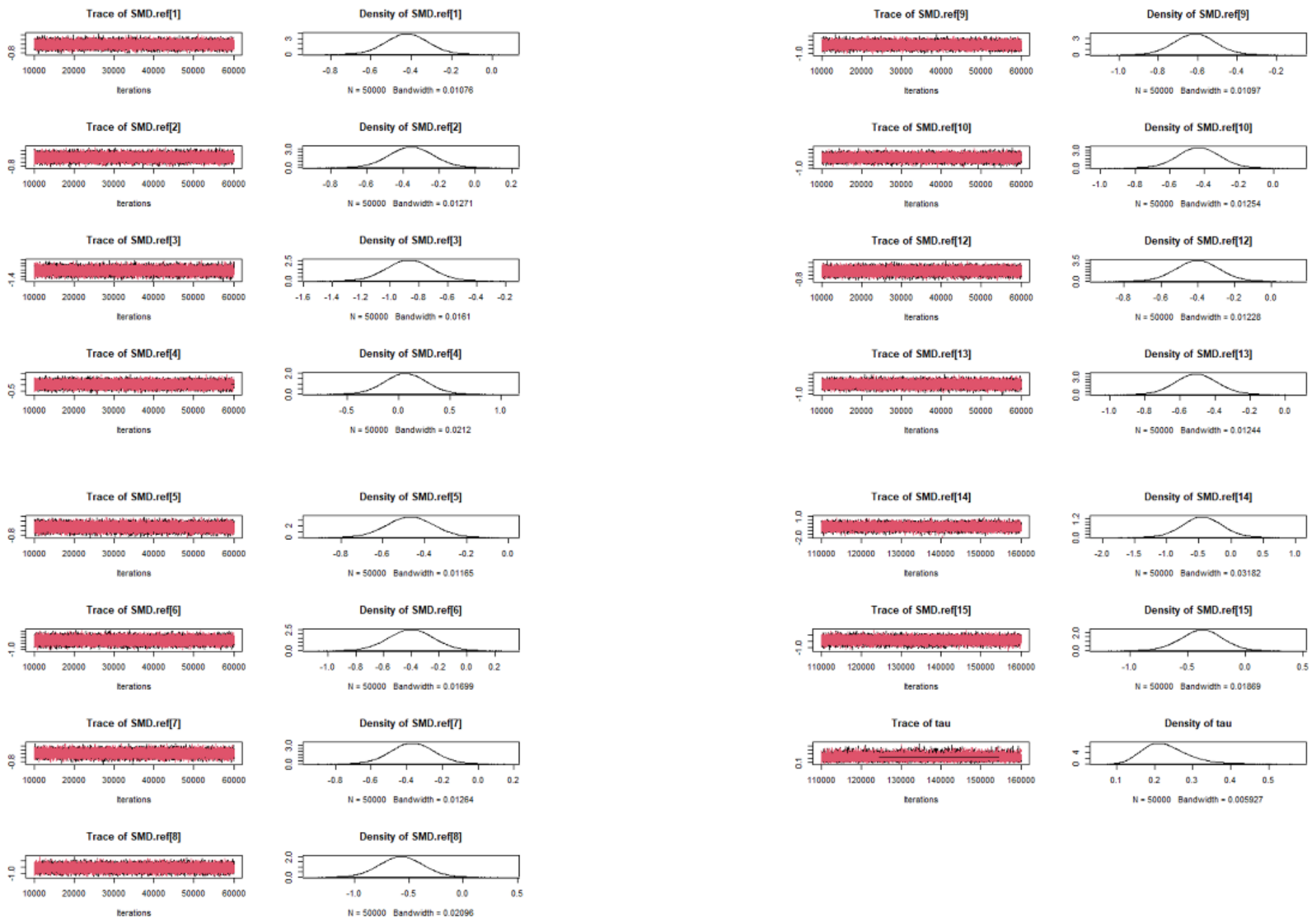

**Appendix Figure 7:** Trace plots for the NMA model with informative priors obtained from GP using a data based approach for $\beta$ and no downweight.

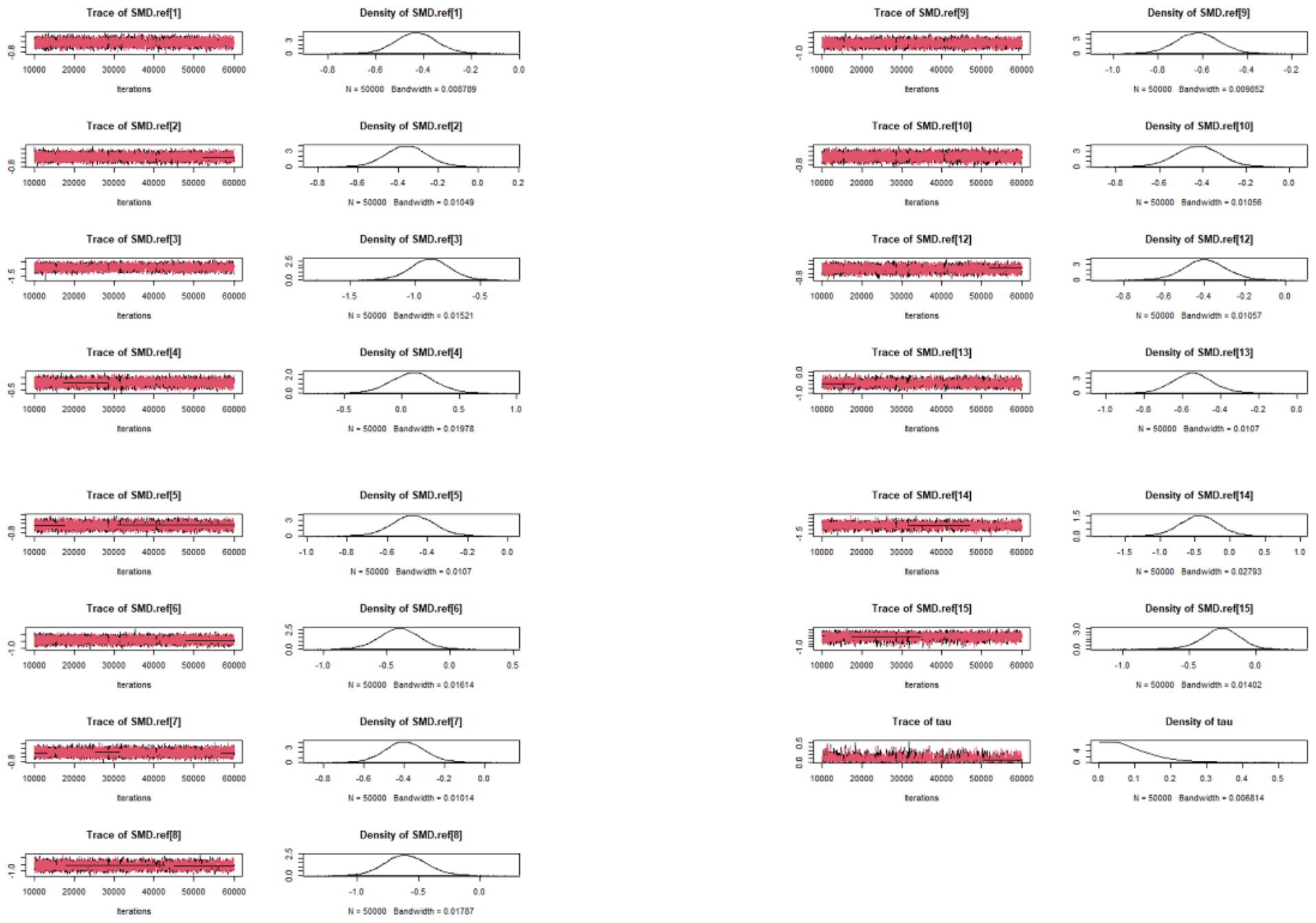

**Appendix Figure 8:** Trace plots for the NMA model with informative priors obtained from GP using a data based approach for $\beta$ and moderate downweight to all GP studies with high RoB

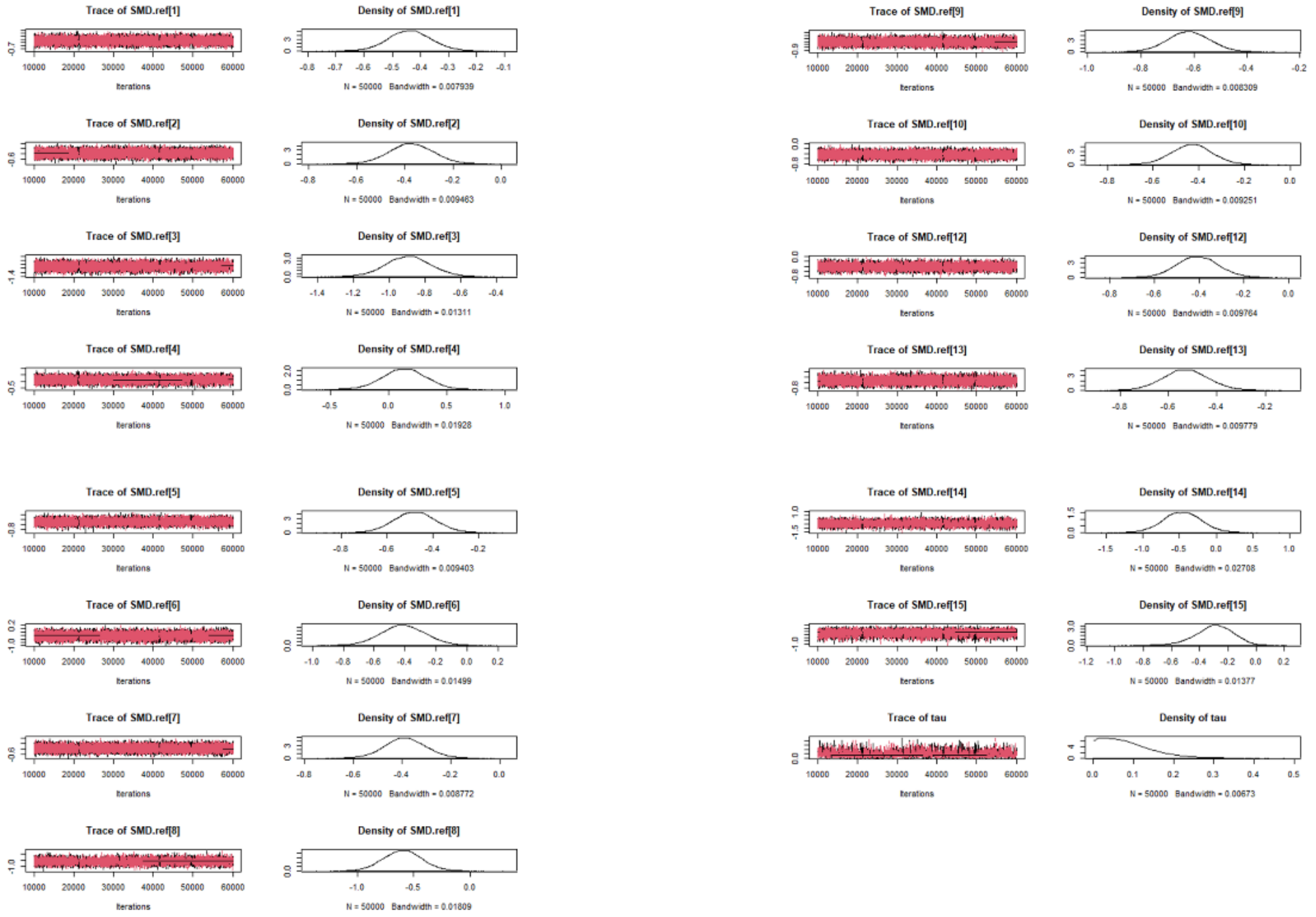

**Appendix Figure 9:** Trace plots for the NMA model with informative priors obtained from GP using a data based approach for $\beta$ and moderate downweight to all GP studies with interventions in $T_a$-$T_c$

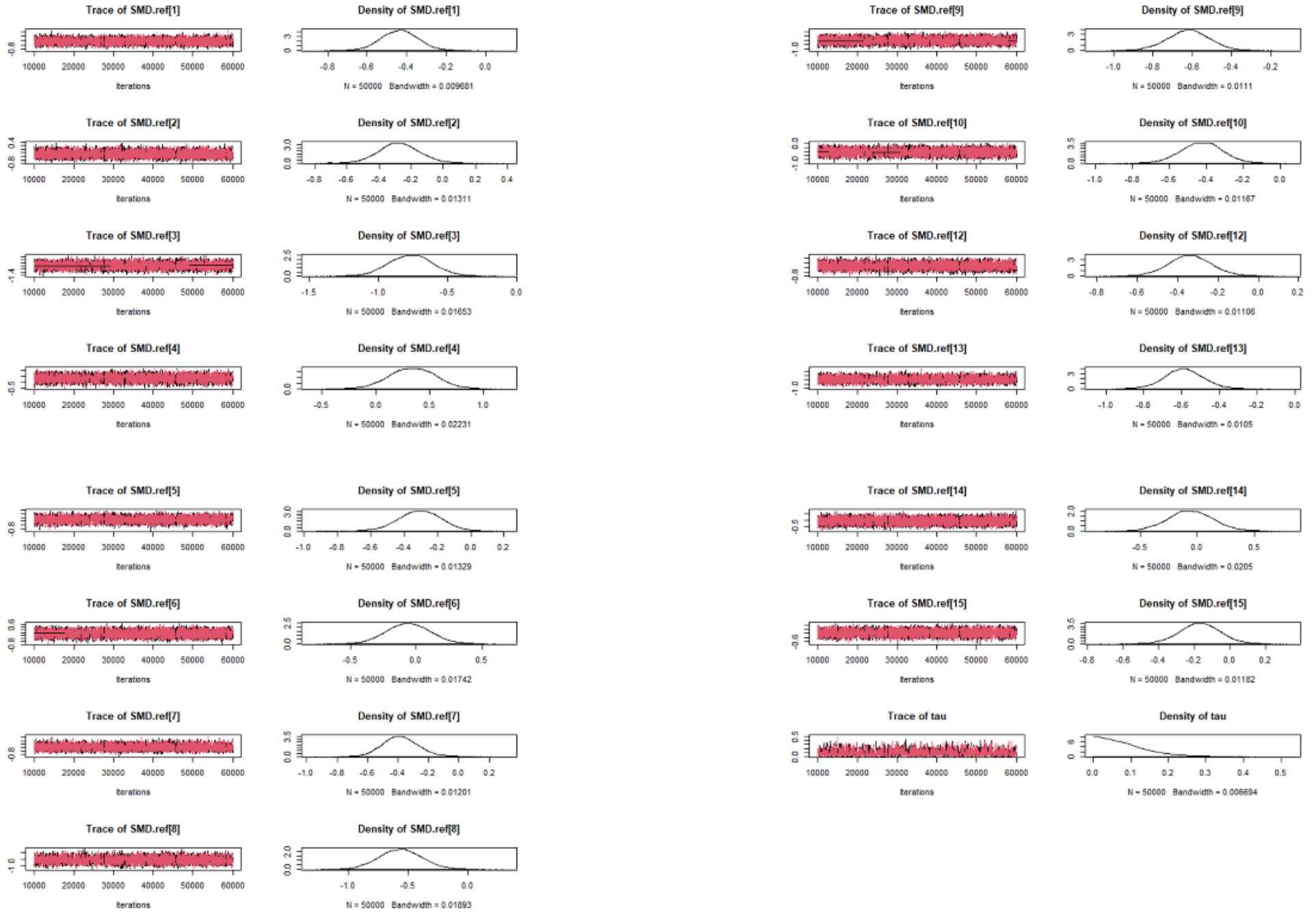

**Appendix Figure 10:** Trace plots for the NMA model with informative priors obtained from GP expert's opinion for $\beta$ and no downweight.

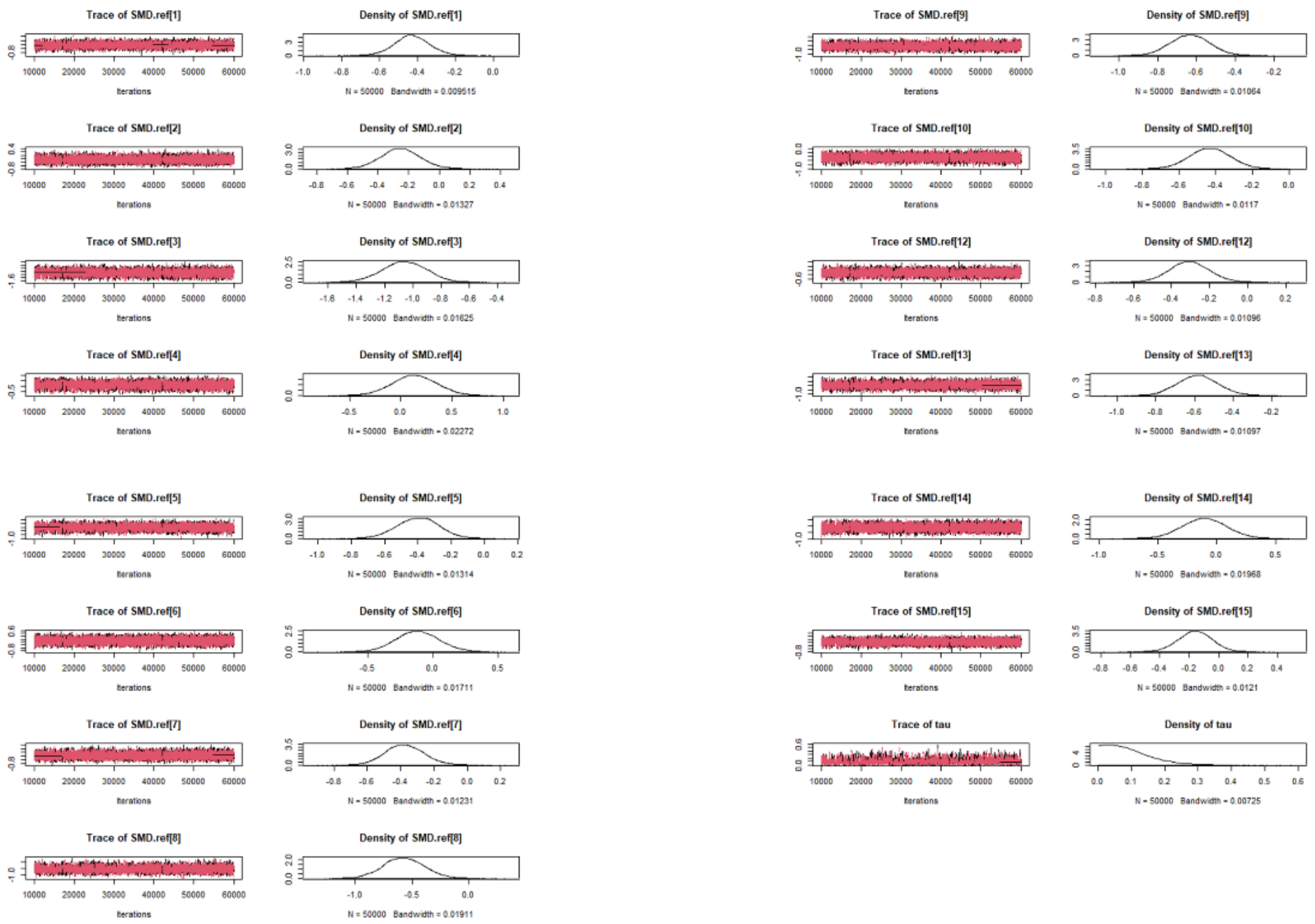

**Appendix Figure 11:** Trace plots for the NMA model with informative priors obtained from GP expert's opinion for $\beta$ and moderate downweight to all GP studies with high RoB

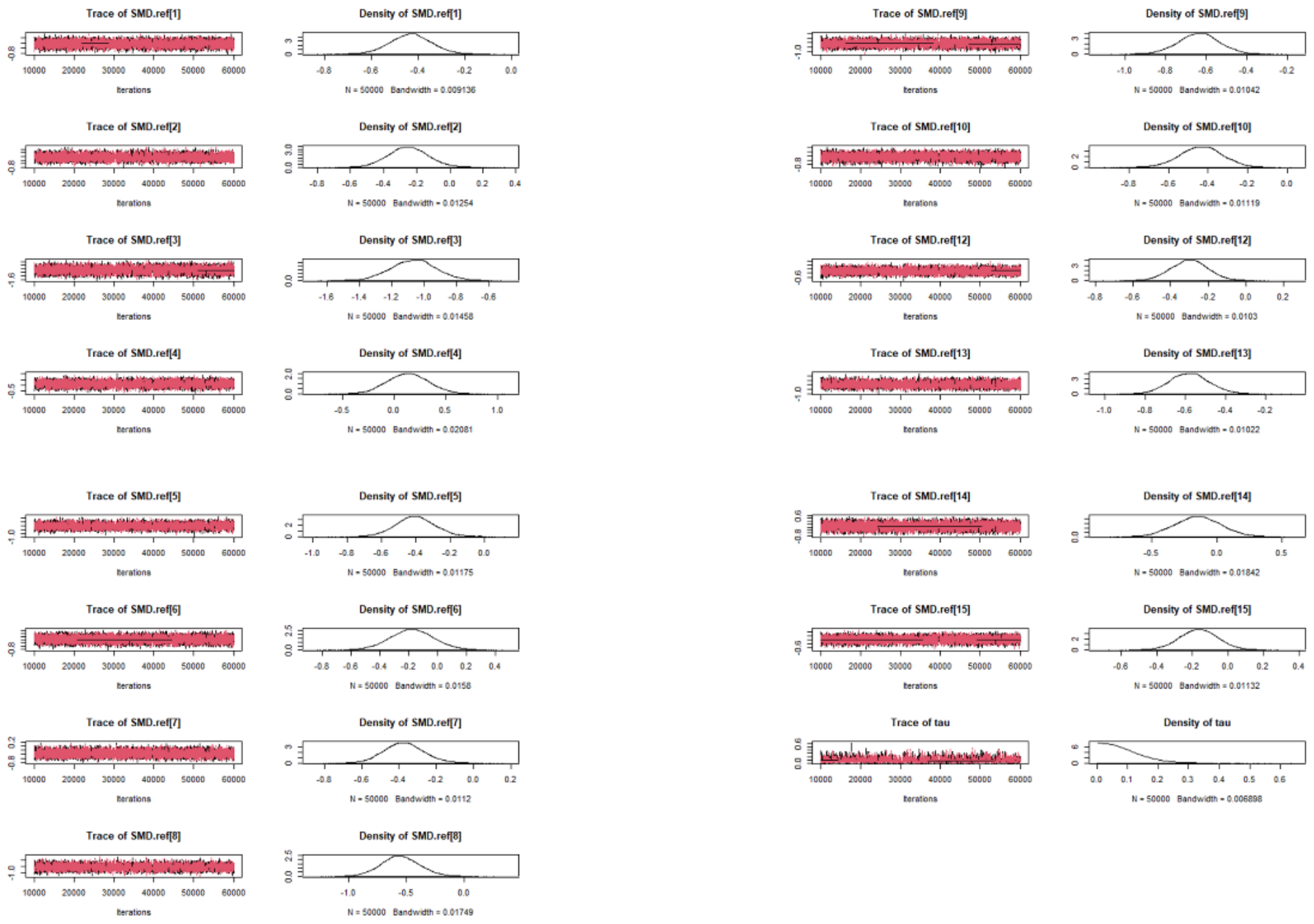

**Appendix Figure 12:** Trace plots for the NMA model with informative priors obtained from GP expert's opinion for $\beta$ and moderate downweight to all GP studies with interventions in $T_a$-$T_c$

# League tables

**Appendix Table 9**: League table for all comparisons in CA network using a naive synthesis NMA model

| | Aripiprazole | Asenapine | Clozapine | Fluphenazine | Haloperidol | Loxapine | Lurasidone | Molindone | Olanzapine | Paliperidone | Placebo | Quetiapine | Risperidone | Trifluoperazine | Ziprasidone |
|---|---|---|---|---|---|---|---|---|---|---|---|---|---|---|---|
| Aripiprazole | NA | -0.065(-0.182,0.057) | 0.441(0.266,0.628) | -0.439(-0.753,-0.134) | 0.02(-0.068,0.107) | -0.115(-0.329,0.093) | -0.088(-0.214,0.03) | 0.128(-0.182,0.446) | 0.148(0.061,0.234) | 0.053(-0.064,0.17) | -0.431(-0.513,-0.352) | -0.019(-0.121,0.089) | 0.092(-0.001,0.183) | -0.158(-0.427,0.127) | -0.074(-0.184,0.036) |
| Asenapine | 0.065(-0.057,0.182) | NA | 0.505(0.318,0.691) | -0.375(-0.689,-0.061) | 0.084(-0.019,0.186) | -0.05(-0.279,0.166) | -0.024(-0.152,0.099) | 0.192(-0.115,0.526) | 0.213(0.116,0.311) | 0.117(-0.016,0.245) | -0.366(-0.458,-0.275) | 0.046(-0.07,0.164) | 0.156(0.05,0.264) | -0.094(-0.373,0.181) | -0.009(-0.132,0.109) |
| Clozapine | -0.441(-0.628,-0.266) | -0.505(-0.691,-0.318) | NA | -0.88(-1.214,-0.534) | -0.421(-0.597,-0.259) | -0.556(-0.815,-0.286) | -0.529(-0.714,-0.345) | -0.313(-0.657,0.025) | -0.292(-0.463,-0.117) | -0.388(-0.574,-0.199) | -0.872(-1.046,-0.703) | -0.459(-0.647,-0.276) | -0.349(-0.526,-0.172) | -0.599(-0.913,-0.277) | -0.515(-0.696,-0.337) |
| Fluphenazine | 0.439(0.134,0.753) | 0.375(0.061,0.689) | 0.88(0.534,1.214) | NA | 0.459(0.156,0.759) | 0.324(-0.04,0.698) | 0.351(0.038,0.666) | 0.567(0.131,0.989) | 0.588(0.288,0.893) | 0.492(0.171,0.807) | 0.008(-0.303,0.307) | 0.421(0.111,0.735) | 0.531(0.221,0.833) | 0.281(-0.127,0.676) | 0.365(0.055,0.674) |
| Haloperidol | -0.02(-0.107,0.068) | -0.084(-0.186,0.019) | 0.421(0.259,0.597) | -0.459(-0.759,-0.156) | NA | -0.135(-0.342,0.072) | -0.108(-0.212,-0.005) | 0.108(-0.192,0.422) | 0.129(0.064,0.197) | 0.033(-0.076,0.138) | -0.45(-0.51,-0.391) | -0.038(-0.123,0.049) | 0.072(0.001,0.139) | -0.178(-0.43,0.09) | -0.093(-0.184,-0.004) |
| Loxapine | 0.115(-0.093,0.329) | 0.05(-0.166,0.279) | 0.556(0.286,0.815) | -0.324(-0.698,0.04) | 0.135(-0.072,0.342) | NA | 0.027(-0.202,0.25) | 0.243(-0.1,0.583) | 0.263(0.056,0.476) | 0.168(-0.045,0.397) | -0.316(-0.518,-0.11) | 0.096(-0.114,0.304) | 0.207(-0.003,0.414) | -0.043(-0.292,0.186) | 0.041(-0.187,0.248) |
| Lurasidone | 0.088(-0.03,0.214) | 0.024(-0.099,0.152) | 0.529(0.345,0.714) | -0.351(-0.666,-0.038) | 0.108(0.005,0.212) | -0.027(-0.25,0.202) | NA | 0.216(-0.091,0.541) | 0.237(0.138,0.338) | 0.141(0.018,0.266) | -0.343(-0.428,-0.256) | 0.07(-0.045,0.182) | 0.18(0.08,0.28) | -0.07(-0.333,0.211) | 0.014(-0.109,0.136) |
| Molindone | -0.128(-0.446,0.182) | -0.192(-0.526,0.115) | 0.313(-0.025,0.657) | -0.567(-0.989,-0.131) | -0.108(-0.422,0.192) | -0.243(-0.583,0.1) | -0.216(-0.541,0.091) | NA | 0.021(-0.286,0.316) | -0.075(-0.407,0.231) | -0.559(-0.874,-0.261) | -0.146(-0.467,0.167) | -0.036(-0.34,0.263) | -0.286(-0.627,0.068) | -0.202(-0.515,0.117) |
| Olanzapine | -0.148(-0.234,-0.061) | -0.213(-0.311,-0.116) | 0.292(0.117,0.463) | -0.588(-0.893,-0.288) | -0.129(-0.197,-0.064) | -0.263(-0.476,-0.056) | -0.237(-0.338,-0.138) | -0.021(-0.316,0.286) | NA | -0.096(-0.196,-0.002) | -0.579(-0.635,-0.525) | -0.167(-0.251,-0.08) | -0.057(-0.123,0.014) | -0.307(-0.563,-0.037) | -0.222(-0.313,-0.136) |
| Paliperidone | -0.053(-0.17,0.064) | -0.117(-0.245,0.016) | 0.388(0.199,0.574) | -0.492(-0.807,-0.171) | -0.033(-0.138,0.076) | -0.168(-0.397,0.045) | -0.141(-0.266,-0.018) | 0.075(-0.231,0.407) | 0.096(0.002,0.196) | NA | -0.484(-0.578,-0.386) | -0.071(-0.19,0.05) | 0.039(-0.066,0.147) | -0.211(-0.497,0.074) | -0.127(-0.246,0.003) |
| Placebo | 0.431(0.352,0.513) | 0.366(0.275,0.458) | 0.872(0.703,1.046) | -0.008(-0.307,0.303) | 0.45(0.391,0.51) | 0.316(0.11,0.518) | 0.343(0.256,0.428) | 0.559(0.261,0.874) | 0.579(0.525,0.635) | 0.484(0.386,0.578) | NA | 0.412(0.337,0.494) | 0.523(0.463,0.585) | 0.273(0.014,0.547) | 0.357(0.267,0.445) |
| Quetiapine | 0.019(-0.089,0.121) | -0.046(-0.164,0.07) | 0.459(0.276,0.647) | -0.421(-0.735,-0.111) | 0.038(-0.049,0.123) | -0.096(-0.304,0.114) | -0.07(-0.182,0.045) | 0.146(-0.167,0.467) | 0.167(0.08,0.251) | 0.071(-0.05,0.19) | -0.412(-0.494,-0.337) | NA | 0.11(0.023,0.197) | -0.14(-0.41,0.137) | -0.055(-0.165,0.054) |
| Risperidone | -0.092(-0.183,0.001) | -0.156(-0.264,-0.05) | 0.349(0.172,0.526) | -0.531(-0.833,-0.221) | -0.072(-0.139,-0.001) | -0.207(-0.414,0.003) | -0.18(-0.28,-0.08) | 0.036(-0.263,0.34) | 0.057(-0.014,0.123) | -0.039(-0.147,0.066) | -0.523(-0.585,-0.463) | -0.11(-0.197,-0.023) | NA | -0.25(-0.51,0.026) | -0.166(-0.26,-0.074) |
| Trifluoperazine | 0.158(-0.127,0.427) | 0.094(-0.181,0.373) | 0.599(0.277,0.913) | -0.281(-0.676,0.127) | 0.178(-0.09,0.43) | 0.043(-0.186,0.292) | 0.07(-0.211,0.333) | 0.286(-0.068,0.627) | 0.307(0.037,0.563) | 0.211(-0.074,0.497) | -0.273(-0.547,-0.014) | 0.14(-0.137,0.41) | 0.25(-0.026,0.51) | NA | 0.084(-0.204,0.352) |
| Ziprasidone | 0.074(-0.036,0.184) | 0.009(-0.109,0.132) | 0.515(0.337,0.696) | -0.365(-0.674,-0.055) | 0.093(0.004,0.184) | -0.041(-0.248,0.187) | -0.014(-0.136,0.109) | 0.202(-0.117,0.515) | 0.222(0.136,0.313) | 0.127(-0.003,0.246) | -0.357(-0.445,-0.267) | 0.055(-0.054,0.165) | 0.166(0.074,0.26) | -0.084(-0.352,0.204) | NA |

**Appendix Table 10:** League table for all comparisons in CA network. Informative priors from GP studies using a data based approach for $\beta$ and no downweight for GP studies.

| | Aripiprazole | Asenapine | Clozapine | Fluphenazine | Haloperidol | Loxapine | Lurasidone | Molindone | Olanzapine | Paliperidone | Placebo | Quetiapine | Risperidone | Trifluoperazine | Ziprasidone |
|---|---|---|---|---|---|---|---|---|---|---|---|---|---|---|---|
| Aripiprazole | NA | -0.07(-0.377,0.258) | 0.446(0.088,0.798) | -0.483(-0.92,-0.046) | 0.036(-0.26,0.338) | -0.026(-0.394,0.342) | -0.052(-0.36,0.247) | 0.144(-0.313,0.592) | 0.185(-0.105,0.481) | 0.011(-0.271,0.291) | -0.425(-0.632,-0.216) | -0.025(-0.306,0.244) | 0.081(-0.256,0.381) | 0.02(-0.607,0.629) | -0.037(-0.438,0.376) |
| Asenapine | 0.07(-0.258,0.377) | NA | 0.517(0.126,0.891) | -0.413(-0.859,0.044) | 0.106(-0.205,0.416) | 0.044(-0.359,0.437) | 0.018(-0.314,0.338) | 0.215(-0.249,0.648) | 0.255(-0.049,0.556) | 0.082(-0.242,0.403) | -0.354(-0.59,-0.129) | 0.045(-0.276,0.354) | 0.151(-0.18,0.472) | 0.09(-0.536,0.696) | 0.033(-0.408,0.446) |
| Clozapine | -0.446(-0.798,-0.088) | -0.517(-0.891,-0.126) | NA | -0.93(-1.397,-0.459) | -0.41(-0.756,-0.059) | -0.472(-0.899,-0.032) | -0.499(-0.86,-0.127) | -0.302(-0.799,0.19) | -0.262(-0.603,0.084) | -0.435(-0.793,-0.06) | -0.871(-1.165,-0.583) | -0.471(-0.837,-0.109) | -0.366(-0.753,0) | -0.427(-1.059,0.234) | -0.484(-0.938,-0.016) |
| Fluphenazine | 0.483(0.046,0.92) | 0.413(-0.044,0.859) | 0.93(0.459,1.397) | NA | 0.519(0.1,0.918) | 0.457(-0.036,0.941) | 0.431(-0.009,0.87) | 0.628(0.055,1.163) | 0.668(0.222,1.087) | 0.495(0.038,0.959) | 0.058(-0.318,0.44) | 0.458(0.018,0.9) | 0.564(0.116,1.023) | 0.503(-0.19,1.161) | 0.446(-0.075,0.979) |
| Haloperidol | -0.036(-0.338,0.26) | -0.106(-0.416,0.205) | 0.41(0.059,0.756) | -0.519(-0.918,-0.1) | NA | -0.062(-0.409,0.281) | -0.088(-0.397,0.233) | 0.108(-0.348,0.554) | 0.149(-0.132,0.437) | -0.025(-0.341,0.28) | -0.461(-0.667,-0.254) | -0.061(-0.374,0.236) | 0.045(-0.273,0.35) | -0.016(-0.626,0.576) | -0.073(-0.477,0.341) |
| Loxapine | 0.026(-0.342,0.394) | -0.044(-0.437,0.359) | 0.472(0.032,0.899) | -0.457(-0.941,0.036) | 0.062(-0.281,0.409) | NA | -0.026(-0.424,0.371) | 0.17(-0.338,0.683) | 0.211(-0.148,0.577) | 0.037(-0.343,0.425) | -0.399(-0.709,-0.092) | 0.001(-0.377,0.384) | 0.107(-0.294,0.504) | 0.046(-0.45,0.546) | -0.011(-0.489,0.482) |
| Lurasidone | 0.052(-0.247,0.36) | -0.018(-0.338,0.314) | 0.499(0.127,0.86) | -0.431(-0.87,0.009) | 0.088(-0.233,0.397) | 0.026(-0.371,0.424) | NA | 0.196(-0.275,0.634) | 0.237(-0.077,0.535) | 0.063(-0.27,0.387) | -0.373(-0.606,-0.146) | 0.027(-0.285,0.353) | 0.133(-0.178,0.45) | 0.072(-0.586,0.679) | 0.015(-0.39,0.432) |
| Molindone | -0.144(-0.592,0.313) | -0.215(-0.648,0.249) | 0.302(-0.19,0.799) | -0.628(-1.163,-0.055) | -0.108(-0.554,0.348) | -0.17(-0.683,0.338) | -0.196(-0.634,0.275) | NA | 0.04(-0.361,0.455) | -0.133(-0.583,0.326) | -0.569(-0.963,-0.183) | -0.169(-0.623,0.276) | -0.064(-0.48,0.375) | -0.125(-0.802,0.572) | -0.182(-0.711,0.333) |
| Olanzapine | -0.185(-0.481,0.105) | -0.255(-0.556,0.049) | 0.262(-0.084,0.603) | -0.668(-1.087,-0.222) | -0.149(-0.437,0.132) | -0.211(-0.577,0.148) | -0.237(-0.535,0.077) | -0.04(-0.455,0.361) | NA | -0.174(-0.474,0.115) | -0.61(-0.811,-0.418) | -0.21(-0.51,0.096) | -0.104(-0.401,0.181) | -0.165(-0.756,0.451) | -0.222(-0.626,0.182) |
| Paliperidone | -0.011(-0.291,0.271) | -0.082(-0.403,0.242) | 0.435(0.06,0.793) | -0.495(-0.959,-0.038) | 0.025(-0.28,0.341) | -0.037(-0.425,0.343) | -0.063(-0.387,0.27) | 0.133(-0.326,0.583) | 0.174(-0.115,0.474) | NA | -0.436(-0.665,-0.204) | -0.036(-0.362,0.283) | 0.07(-0.249,0.403) | 0.008(-0.621,0.633) | -0.049(-0.479,0.376) |
| Placebo | 0.425(0.216,0.632) | 0.354(0.129,0.59) | 0.871(0.583,1.165) | -0.058(-0.44,0.318) | 0.461(0.254,0.667) | 0.399(0.092,0.709) | 0.373(0.146,0.606) | 0.569(0.183,0.963) | 0.61(0.418,0.811) | 0.436(0.204,0.665) | NA | 0.4(0.168, 0.621) | 0.506(0.277,0.738) | 0.444(-0.125,1.002) | 0.387(0.044, 0.735) |
| Quetiapine | 0.025(-0.244,0.306) | -0.045(-0.354,0.276) | 0.471(0.109,0.837) | -0.458(-0.9,-0.018) | 0.061(-0.236,0.374) | -0.001(-0.384,0.377) | -0.027(-0.353,0.285) | 0.169(-0.276,0.623) | 0.21(-0.096,0.51) | 0.036(-0.283,0.362) | -0.4(-0.621,-0.168) | NA | 0.106(-0.197,0.412) | 0.045(-0.583,0.647) | -0.012(-0.426,0.402) |
| Risperidone | -0.081(-0.381,0.256) | -0.151(-0.472,0.18) | 0.366(0,0.753) | -0.564(-1.023,-0.116) | -0.045(-0.35,0.273) | -0.107(-0.504,0.294) | -0.133(-0.45,0.178) | 0.064(-0.375,0.48) | 0.104(-0.181,0.401) | -0.07(-0.403,0.249) | -0.506(-0.738,-0.277) | -0.106(-0.412,0.197) | NA | -0.061(-0.668,0.57) | -0.118(-0.539,0.297) |
| Trifluoperazine | -0.02(-0.629,0.607) | -0.09(-0.696,0.536) | 0.427(-0.234,1.059) | -0.503(-1.161,0.19) | 0.016(-0.576,0.626) | -0.046(-0.546,0.45) | -0.072(-0.679,0.586) | 0.125(-0.572,0.802) | 0.165(-0.451,0.756) | -0.008(-0.633,0.621) | -0.444(-1.002,0.125) | -0.045(-0.647,0.583) | 0.061(-0.57,0.668) | NA | -0.057(-0.721,0.626) |
| Ziprasidone | 0.037(-0.376,0.438) | -0.033(-0.446,0.408) | 0.484(0.016,0.938) | -0.446(-0.979,0.075) | 0.073(-0.341,0.477) | 0.011(-0.482,0.489) | -0.015(-0.432,0.39) | 0.182(-0.333,0.711) | 0.222(-0.182,0.626) | 0.049(-0.376,0.479) | -0.387(-0.735,-0.044) | 0.012(-0.402,0.426) | 0.118(-0.297,0.539) | 0.057(-0.626,0.721) | NA |

**Appendix Table 11:** League table for all comparisons in CA network. Informative priors from GP studies using a data based approach for $\beta$ and moderate downweight for GP studies with high RoB.

| | Aripiprazole | Asenapine | Clozapine | Fluphenazine | Haloperidol | Loxapine | Lurasidone | Molindone | Olanzapine | Paliperidone | Placebo | Quetiapine | Risperidone | Trifluoperazine | Ziprasidone |
|---|---|---|---|---|---|---|---|---|---|---|---|---|---|---|---|
| Aripiprazole | NA | -0.065(-0.314,0.182) | 0.457(0.122,0.767) | -0.536(-0.927,-0.128) | 0.047(-0.216,0.312) | -0.018(-0.358,0.311) | -0.026(-0.271,0.224) | 0.195(-0.186,0.57) | 0.204(-0.03,0.45) | -0.002(-0.216,0.213) | -0.423(-0.59,-0.262) | -0.027(-0.262,0.215) | 0.125(-0.13,0.376) | 0.023(-0.522,0.572) | -0.141(-0.447,0.171) |
| Asenapine | 0.065(-0.182,0.314) | NA | 0.522(0.171,0.888) | -0.47(-0.88,-0.048) | 0.113(-0.162,0.383) | 0.047(-0.302,0.383) | 0.039(-0.231,0.304) | 0.26(-0.119,0.657) | 0.269(-0.002,0.53) | 0.063(-0.217,0.336) | -0.357(-0.547,-0.154) | 0.038(-0.227,0.31) | 0.19(-0.099,0.465) | 0.089(-0.455,0.628) | -0.076(-0.392,0.267) |
| Clozapine | -0.457(-0.767,-0.122) | -0.522(-0.888,-0.171) | NA | -0.993(-1.467,-0.515) | -0.41(-0.754,-0.085) | -0.475(-0.893,-0.063) | -0.483(-0.828,-0.151) | -0.262(-0.697,0.156) | -0.253(-0.58,0.072) | -0.459(-0.809,-0.11) | -0.88(-1.164,-0.598) | -0.484(-0.826,-0.138) | -0.332(-0.686,0.011) | -0.434(-1.052,0.179) | -0.598(-0.986,-0.193) |
| Fluphenazine | 0.536(0.128,0.927) | 0.47(0.048,0.88) | 0.993(0.515,1.467) | NA | 0.583(0.189,0.961) | 0.517(0.06,0.963) | 0.509(0.063,0.926) | 0.73(0.207,1.262) | 0.739(0.306,1.14) | 0.533(0.108,0.948) | 0.113(-0.258,0.472) | 0.508(0.092,0.919) | 0.661(0.237,1.07) | 0.559(-0.07,1.147) | 0.395(-0.042,0.851) |
| Haloperidol | -0.047(-0.312,0.216) | -0.113(-0.383,0.162) | 0.41(0.085,0.754) | -0.583(-0.961,-0.189) | NA | -0.065(-0.394,0.275) | -0.074(-0.351,0.201) | 0.147(-0.25,0.581) | 0.156(-0.095,0.417) | -0.05(-0.329,0.252) | -0.47(-0.676,-0.27) | -0.075(-0.348,0.204) | 0.078(-0.202,0.352) | -0.024(-0.561,0.521) | -0.188(-0.525,0.165) |
| Loxapine | 0.018(-0.311,0.358) | -0.047(-0.383,0.302) | 0.475(0.063,0.893) | -0.517(-0.963,-0.06) | 0.065(-0.275,0.394) | NA | -0.008(-0.373,0.338) | 0.213(-0.223,0.66) | 0.222(-0.124,0.57) | 0.016(-0.329,0.371) | -0.405(-0.703,-0.113) | -0.009(-0.365,0.369) | 0.143(-0.219,0.501) | 0.041(-0.358,0.455) | -0.123(-0.524,0.277) |
| Lurasidone | 0.026(-0.224,0.271) | -0.039(-0.304,0.231) | 0.483(0.151,0.828) | -0.509(-0.926,-0.063) | 0.074(-0.201,0.351) | 0.008(-0.338,0.373) | NA | 0.221(-0.168,0.579) | 0.23(-0.035,0.497) | 0.024(-0.244,0.303) | -0.396(-0.588,-0.201) | -0.001(-0.273,0.278) | 0.151(-0.123,0.422) | 0.05(-0.506,0.595) | -0.115(-0.429,0.236) |
| Molindone | -0.195(-0.57,0.186) | -0.26(-0.657,0.119) | 0.262(-0.156,0.697) | -0.73(-1.262,-0.207) | -0.147(-0.581,0.25) | -0.213(-0.66,0.223) | -0.221(-0.579,0.168) | NA | 0.009(-0.323,0.366) | -0.197(-0.586,0.204) | -0.617(-0.951,-0.283) | -0.222(-0.621,0.177) | -0.069(-0.412,0.282) | -0.171(-0.789,0.441) | -0.336(-0.753,0.106) |
| Olanzapine | -0.204(-0.45,0.03) | -0.269(-0.53,0.002) | 0.253(-0.072,0.58) | -0.739(-1.14,-0.306) | -0.156(-0.417,0.095) | -0.222(-0.57,0.124) | -0.23(-0.497,0.035) | -0.009(-0.366,0.323) | NA | -0.206(-0.496,0.061) | -0.626(-0.809,-0.44) | -0.231(-0.494,0.046) | -0.079(-0.313,0.152) | -0.18(-0.729,0.362) | -0.345(-0.664,-0.009) |
| Paliperidone | 0.002(-0.213,0.216) | -0.063(-0.336,0.217) | 0.459(0.11,0.809) | -0.533(-0.948,-0.108) | 0.05(-0.252,0.329) | -0.016(-0.371,0.329) | -0.024(-0.303,0.244) | 0.197(-0.204,0.586) | 0.206(-0.061,0.496) | NA | -0.42(-0.614,-0.236) | -0.025(-0.311,0.239) | 0.127(-0.171,0.386) | 0.026(-0.536,0.58) | -0.139(-0.455,0.197) |
| Placebo | 0.423(0.262,0.59) | 0.357(0.154,0.547) | 0.88(0.598,1.164) | -0.113(-0.472,0.258) | 0.47(0.27,0.676) | 0.405(0.113,0.703) | 0.396(0.201,0.588) | 0.617(0.283,0.951) | 0.626(0.44,0.809) | 0.42(0.236,0.614) | NA | 0.395(0.191,0.6) | 0.548(0.343,0.737) | 0.446(-0.082,0.976) | 0.282(0.03,0.567) |
| Quetiapine | 0.027(-0.215,0.262) | -0.038(-0.31,0.227) | 0.484(0.138,0.826) | -0.508(-0.919,-0.092) | 0.075(-0.204,0.348) | 0.009(-0.369,0.365) | 0.001(-0.278,0.273) | 0.222(-0.177,0.621) | 0.231(-0.046,0.494) | 0.025(-0.239,0.311) | -0.395(-0.6,-0.191) | NA | 0.152(-0.124,0.424) | 0.051(-0.509,0.62) | -0.114(-0.425,0.229) |
| Risperidone | -0.125(-0.376,0.13) | -0.19(-0.465,0.099) | 0.332(-0.011,0.686) | -0.661(-1.07,-0.237) | -0.078(-0.352,0.202) | -0.143(-0.501,0.219) | -0.151(-0.422,0.123) | 0.069(-0.282,0.412) | 0.079(-0.152,0.313) | -0.127(-0.386,0.171) | -0.548(-0.737,-0.343) | -0.152(-0.424,0.124) | NA | -0.102(-0.651,0.483) | -0.266(-0.592,0.085) |
| Trifluoperazine | -0.023(-0.572,0.522) | -0.089(-0.628,0.455) | 0.434(-0.179,1.052) | -0.559(-1.147,0.07) | 0.024(-0.521,0.561) | -0.041(-0.455,0.358) | -0.05(-0.595,0.506) | 0.171(-0.441,0.789) | 0.18(-0.362,0.729) | -0.026(-0.58,0.536) | -0.446(-0.976,0.082) | -0.051(-0.62,0.509) | 0.102(-0.483,0.651) | NA | -0.164(-0.752,0.433) |
| Ziprasidone | 0.141(-0.171,0.447) | 0.076(-0.267,0.392) | 0.598(0.193,0.986) | -0.395(-0.851,0.042) | 0.188(-0.165,0.525) | 0.123(-0.277,0.524) | 0.115(-0.236,0.429) | 0.336(-0.106,0.753) | 0.345(0.009,0.664) | 0.139(-0.197,0.455) | -0.282(-0.567,-0.03) | 0.114(-0.229,0.425) | 0.266(-0.085,0.592) | 0.164(-0.433,0.752) | NA |

**Appendix Table 12:** League table for all comparisons in CA network. Informative priors from GP studies using a data based approach for $\beta$ and moderate downweight for GP studies evaluating interventions in $T_{\alpha} - T_{c}$.

| | Aripiprazole | Asenapine | Clozapine | Fluphenazine | Haloperidol | Loxapine | Lurasidone | Molindone | Olanzapine | Paliperidone | Placebo | Quetiapine | Risperidone | Trifluoperazine | Ziprasidone |
|---|---|---|---|---|---|---|---|---|---|---|---|---|---|---|---|
| Aripiprazole | NA | -0.066(-0.325,0.188) | 0.447(0.149,0.766) | -0.503(-0.895,-0.12) | 0.037(-0.208,0.28) | -0.023(-0.355,0.302) | -0.078(-0.333,0.169) | 0.113(-0.312,0.528) | 0.166(-0.077,0.396) | 0.01(-0.217,0.246) | -0.439(-0.613,-0.269) | -0.016(-0.271,0.238) | 0.045(-0.224,0.318) | 0.021(-0.567,0.628) | -0.035(-0.411,0.356) |
| Asenapine | 0.066(-0.188,0.325) | NA | 0.513(0.203,0.827) | -0.437(-0.844,-0.019) | 0.103(-0.146,0.352) | 0.043(-0.293,0.379) | -0.013(-0.277,0.249) | 0.178(-0.269,0.596) | 0.232(-0.006,0.485) | 0.075(-0.188,0.35) | -0.373(-0.564,-0.183) | 0.049(-0.229,0.334) | 0.111(-0.167,0.385) | 0.087(-0.507,0.69) | 0.03(-0.346,0.414) |
| Clozapine | -0.447(-0.766,-0.149) | -0.513(-0.827,-0.203) | NA | -0.95(-1.398,-0.492) | -0.41(-0.707,-0.103) | -0.47(-0.857,-0.107) | -0.526(-0.849,-0.216) | -0.334(-0.8,0.101) | -0.281(-0.568,0.012) | -0.438(-0.753,-0.122) | -0.886(-1.14,-0.632) | -0.463(-0.783,-0.14) | -0.402(-0.726,-0.084) | -0.426(-1.053,0.19) | -0.482(-0.895,-0.03) |
| Fluphenazine | 0.503(0.12,0.895) | 0.437(0.019,0.844) | 0.95(0.492,1.398) | NA | 0.54(0.147,0.929) | 0.48(0.021,0.936) | 0.424(0.008,0.82) | 0.616(0.09,1.113) | 0.669(0.286,1.049) | 0.513(0.108,0.916) | 0.064(-0.297,0.428) | 0.487(0.078,0.885) | 0.548(0.123,0.959) | 0.524(-0.162,1.195) | 0.468(-0.016,0.965) |
| Haloperidol | -0.037(-0.28,0.208) | -0.103(-0.352,0.146) | 0.41(0.103,0.707) | -0.54(-0.929,-0.147) | NA | -0.06(-0.386,0.264) | -0.116(-0.371,0.146) | 0.076(-0.352,0.48) | 0.129(-0.091,0.357) | -0.027(-0.263,0.221) | -0.476(-0.644,-0.31) | -0.053(-0.319,0.205) | 0.008(-0.258,0.266) | -0.016(-0.617,0.551) | -0.072(-0.44,0.315) |
| Loxapine | 0.023(-0.302,0.355) | -0.043(-0.379,0.293) | 0.47(0.107,0.857) | -0.48(-0.936,-0.021) | 0.06(-0.264,0.386) | NA | -0.056(-0.409,0.286) | 0.135(-0.364,0.586) | 0.188(-0.128,0.522) | 0.032(-0.3,0.383) | -0.416(-0.695,-0.13) | 0.006(-0.351,0.367) | 0.068(-0.283,0.423) | 0.044(-0.439,0.569) | -0.013(-0.455,0.451) |
| Lurasidone | 0.078(-0.169,0.333) | 0.013(-0.249,0.277) | 0.526(0.216,0.849) | -0.424(-0.82,-0.008) | 0.116(-0.146,0.371) | 0.056(-0.286,0.409) | NA | 0.191(-0.248,0.599) | 0.244(0.005,0.488) | 0.088(-0.178,0.35) | -0.36(-0.535,-0.176) | 0.062(-0.201,0.336) | 0.123(-0.153,0.39) | 0.1(-0.5,0.712) | 0.043(-0.333,0.427) |
| Molindone | -0.113(-0.528,0.312) | -0.178(-0.596,0.269) | 0.334(-0.101,0.8) | -0.616(-1.113,-0.09) | -0.076(-0.48,0.352) | -0.135(-0.586,0.364) | -0.191(-0.599,0.248) | NA | 0.053(-0.35,0.459) | -0.103(-0.516,0.325) | -0.551(-0.924,-0.139) | -0.129(-0.562,0.324) | -0.068(-0.473,0.349) | -0.092(-0.746,0.646) | -0.148(-0.635,0.378) |
| Olanzapine | -0.166(-0.396,0.077) | -0.232(-0.485,0.006) | 0.281(-0.012,0.568) | -0.669(-1.049,-0.286) | -0.129(-0.357,0.091) | -0.188(-0.522,0.128) | -0.244(-0.488,-0.005) | -0.053(-0.459,0.35) | NA | -0.156(-0.395,0.086) | -0.605(-0.766,-0.447) | -0.182(-0.44,0.074) | -0.121(-0.361,0.125) | -0.145(-0.737,0.443) | -0.201(-0.58,0.169) |
| Paliperidone | -0.01(-0.246,0.217) | -0.075(-0.35,0.188) | 0.438(0.122,0.753) | -0.513(-0.916,-0.108) | 0.027(-0.221,0.263) | -0.032(-0.383,0.3) | -0.088(-0.35,0.178) | 0.103(-0.325,0.516) | 0.156(-0.086,0.395) | NA | -0.448(-0.638,-0.263) | -0.026(-0.285,0.247) | 0.035(-0.243,0.311) | 0.012(-0.596,0.622) | -0.045(-0.403,0.359) |
| Placebo | 0.439(0.269,0.613) | 0.373(0.183,0.564) | 0.886(0.632,1.14) | -0.064(-0.428,0.297) | 0.476(0.31,0.644) | 0.416(0.13,0.695) | 0.36(0.176,0.535) | 0.551(0.139,0.924) | 0.605(0.447,0.766) | 0.448(0.263,0.638) | NA | 0.422(0.223,0.627) | 0.484(0.274,0.687) | 0.46(-0.113,1.028) | 0.403(0.077,0.752) |
| Quetiapine | 0.016(-0.238,0.271) | -0.049(-0.334,0.229) | 0.463(0.14,0.783) | -0.487(-0.885,-0.078) | 0.053(-0.205,0.319) | -0.006(-0.367,0.351) | -0.062(-0.336,0.201) | 0.129(-0.324,0.562) | 0.182(-0.074,0.44) | 0.026(-0.247,0.285) | -0.422(-0.627,-0.223) | NA | 0.061(-0.237,0.341) | 0.037(-0.57,0.632) | -0.019(-0.383,0.365) |
| Risperidone | -0.045(-0.318,0.224) | -0.111(-0.385,0.167) | 0.402(0.084,0.726) | -0.548(-0.959,-0.123) | -0.008(-0.266,0.258) | -0.068(-0.423,0.283) | -0.123(-0.39,0.153) | 0.068(-0.349,0.473) | 0.121(-0.125,0.361) | -0.035(-0.311,0.243) | -0.484(-0.687,-0.274) | -0.061(-0.341,0.237) | NA | -0.024(-0.626,0.592) | -0.08(-0.472,0.341) |
| Trifluoperazine | -0.021(-0.628,0.567) | -0.087(-0.69,0.507) | 0.426(-0.19,1.053) | -0.524(-1.195,0.162) | 0.016(-0.551,0.617) | -0.044(-0.569,0.439) | -0.1(-0.712,0.5) | 0.092(-0.646,0.746) | 0.145(-0.443,0.737) | -0.012(-0.622,0.596) | -0.46(-1.028,0.113) | -0.037(-0.632,0.57) | 0.024(-0.592,0.626) | NA | -0.056(-0.69,0.64) |
| Ziprasidone | 0.035(-0.356,0.411) | -0.03(-0.414,0.346) | 0.482(0.03,0.895) | -0.468(-0.965,0.016) | 0.072(-0.315,0.44) | 0.013(-0.451,0.455) | -0.043(-0.427,0.333) | 0.148(-0.378,0.635) | 0.201(-0.169,0.58) | 0.045(-0.359,0.403) | -0.403(-0.752,-0.077) | 0.019(-0.365,0.383) | 0.08(-0.341,0.472) | 0.056(-0.64,0.69) | NA |

**Appendix Table 13:** League table for all comparisons in CA network. Informative priors from GP studies using expert's opinion for $\beta$ and no downweight for GP studies.

| | Aripiprazole | Asenapine | Clozapine | Fluphenazine | Haloperidol | Loxapine | Lurasidone | Molindone | Olanzapine | Paliperidone | Placebo | Quetiapine | Risperidone | Trifluoperazine | Ziprasidone |
|---|---|---|---|---|---|---|---|---|---|---|---|---|---|---|---|
| Aripiprazole | NA | -0.162(-0.475,0.128) | 0.334(-0.017,0.682) | -0.799(-1.272,-0.347) | -0.132(-0.436,0.157) | -0.375(-0.739,-0.032) | -0.052(-0.356,0.224) | 0.134(-0.249,0.493) | 0.183(-0.094,0.463) | -0.013(-0.257,0.228) | -0.435(-0.617,-0.259) | -0.092(-0.342,0.156) | 0.157(-0.123,0.432) | -0.382(-0.799,0.032) | -0.268(-0.563,0.023) |
| Asenapine | 0.162(-0.128,0.475) | NA | 0.496(0.107,0.871) | -0.637(-1.131,-0.133) | 0.03(-0.306,0.37) | -0.213(-0.589,0.171) | 0.11(-0.213,0.43) | 0.296(-0.12,0.721) | 0.344(0.041,0.662) | 0.149(-0.159,0.49) | -0.274(-0.513,-0.024) | 0.07(-0.237,0.389) | 0.318(-0.015,0.62) | -0.22(-0.676,0.229) | -0.106(-0.43,0.219) |
| Clozapine | -0.334(-0.682,0.017) | -0.496(-0.871,-0.107) | NA | -1.133(-1.654,-0.623) | -0.466(-0.818,-0.094) | -0.71(-1.134,-0.283) | -0.386(-0.755,0) | -0.2(-0.628,0.255) | -0.152(-0.472,0.193) | -0.347(-0.709,0.015) | -0.77(-1.068,-0.469) | -0.426(-0.785,-0.057) | -0.178(-0.522,0.195) | -0.716(-1.194,-0.257) | -0.602(-0.996,-0.237) |
| Fluphenazine | 0.799(0.347,1.272) | 0.637(0.133,1.131) | 1.133(0.623,1.654) | NA | 0.667(0.226,1.098) | 0.424(-0.105,0.952) | 0.747(0.239,1.236) | 0.933(0.391,1.477) | 0.981(0.492,1.466) | 0.786(0.314,1.264) | 0.363(-0.054,0.785) | 0.707(0.232,1.177) | 0.955(0.474,1.433) | 0.417(-0.157,0.994) | 0.531(0.032,1.038) |
| Haloperidol | 0.132(-0.157,0.436) | -0.03(-0.37,0.306) | 0.466(0.094,0.818) | -0.667(-1.098,-0.226) | NA | -0.243(-0.599,0.134) | 0.08(-0.235,0.414) | 0.267(-0.161,0.676) | 0.315(0.005,0.612) | 0.119(-0.199,0.439) | -0.303(-0.541,-0.055) | 0.04(-0.274,0.349) | 0.289(-0.022,0.599) | -0.25(-0.671,0.164) | -0.136(-0.456,0.201) |
| Loxapine | 0.375(0.032,0.739) | 0.213(-0.171,0.589) | 0.71(0.283,1.134) | -0.424(-0.952,0.105) | 0.243(-0.134,0.599) | NA | 0.324(-0.048,0.689) | 0.51(0.023,0.968) | 0.558(0.171,0.914) | 0.362(-0.013,0.753) | -0.06(-0.38,0.256) | 0.283(-0.08,0.649) | 0.532(0.166,0.905) | -0.006(-0.354,0.33) | 0.107(-0.266,0.508) |
| Lurasidone | 0.052(-0.224,0.356) | -0.11(-0.43,0.213) | 0.386(0,0.755) | -0.747(-1.236,-0.239) | -0.08(-0.414,0.235) | -0.324(-0.689,0.048) | NA | 0.186(-0.221,0.597) | 0.234(-0.053,0.537) | 0.039(-0.283,0.362) | -0.384(-0.598,-0.152) | -0.04(-0.331,0.274) | 0.208(-0.084,0.515) | -0.33(-0.76,0.106) | -0.216(-0.538,0.105) |
| Molindone | -0.134(-0.493,0.249) | -0.296(-0.721,0.12) | 0.2(-0.255,0.628) | -0.933(-1.477,-0.391) | -0.267(-0.676,0.161) | -0.51(-0.968,-0.023) | -0.186(-0.597,0.221) | NA | 0.048(-0.278,0.389) | -0.148(-0.533,0.261) | -0.57(-0.906,-0.224) | -0.226(-0.593,0.161) | 0.022(-0.34,0.389) | -0.516(-1.04,-0.006) | -0.402(-0.809,0.003) |
| Olanzapine | -0.183(-0.463,0.094) | -0.344(-0.662,-0.041) | 0.152(-0.193,0.472) | -0.981(-1.466,-0.492) | -0.315(-0.612,-0.005) | -0.558(-0.914,-0.171) | -0.234(-0.537,0.053) | -0.048(-0.389,0.278) | NA | -0.196(-0.498,0.105) | -0.618(-0.824,-0.414) | -0.274(-0.558,-0.004) | -0.026(-0.286,0.235) | -0.564(-0.992,-0.136) | -0.45(-0.77,-0.144) |
| Paliperidone | 0.013(-0.228,0.257) | -0.149(-0.49,0.159) | 0.347(-0.015,0.709) | -0.786(-1.264,-0.314) | -0.119(-0.439,0.199) | -0.362(-0.753,0.013) | -0.039(-0.362,0.283) | 0.148(-0.261,0.533) | 0.196(-0.105,0.498) | NA | -0.422(-0.634,-0.197) | -0.079(-0.387,0.219) | 0.17(-0.141,0.457) | -0.369(-0.808,0.069) | -0.255(-0.572,0.062) |
| Placebo | 0.435(0.259,0.617) | 0.274(0.024,0.513) | 0.77(0.469,1.068) | -0.363(-0.785,0.054) | 0.303(0.055,0.541) | 0.06(-0.256,0.38) | 0.384(0.152,0.598) | 0.57(0.224,0.906) | 0.618(0.414,0.824) | 0.422(0.197,0.634) | NA | 0.344(0.138,0.542) | 0.592(0.379,0.794) | 0.054(-0.318,0.431) | 0.168(-0.057,0.389) |
| Quetiapine | 0.092(-0.156,0.342) | -0.07(-0.389,0.237) | 0.426(0.057,0.785) | -0.707(-1.177,-0.232) | -0.04(-0.349,0.274) | -0.283(-0.649,0.08) | 0.04(-0.274,0.331) | 0.226(-0.161,0.593) | 0.274(0.004,0.558) | 0.079(-0.219,0.387) | -0.344(-0.542,-0.138) | NA | 0.248(-0.027,0.51) | -0.29(-0.714,0.129) | -0.176(-0.481,0.128) |
| Risperidone | -0.157(-0.432,0.123) | -0.318(-0.62,0.015) | 0.178(-0.195,0.522) | -0.955(-1.433,-0.474) | -0.289(-0.599,0.022) | -0.532(-0.905,-0.166) | -0.208(-0.515,0.084) | -0.022(-0.389,0.34) | 0.026(-0.235,0.286) | -0.17(-0.457,0.141) | -0.592(-0.794,-0.379) | -0.248(-0.51,0.027) | NA | -0.538(-0.945,-0.108) | -0.424(-0.733,-0.117) |
| Trifluoperazine | 0.382(-0.032,0.799) | 0.22(-0.229,0.676) | 0.716(0.257,1.194) | -0.417(-0.994,0.157) | 0.25(-0.164,0.671) | 0.006(-0.33,0.354) | 0.33(-0.106,0.76) | 0.516(0.006,1.04) | 0.564(0.136,0.992) | 0.369(-0.069,0.808) | -0.054(-0.431,0.318) | 0.29(-0.129,0.714) | 0.538(0.108,0.945) | NA | 0.114(-0.313,0.56) |
| Ziprasidone | 0.268(-0.023,0.563) | 0.106(-0.219,0.43) | 0.602(0.237,0.996) | -0.531(-1.038,-0.032) | 0.136(-0.201,0.456) | -0.107(-0.508,0.266) | 0.216(-0.105,0.538) | 0.402(-0.003,0.809) | 0.45(0.144,0.77) | 0.255(-0.062,0.572) | -0.168(-0.389,0.057) | 0.176(-0.128,0.481) | 0.424(0.117,0.733) | -0.114(-0.56,0.313) | NA |

**Appendix Table 14:** League table for all comparisons in CA network. Informative priors from GP studies using expert's opinion for $\beta$ and moderate downweight for GP studies with high RoB

| | Aripiprazole | Asenapine | Clozapine | Fluphenazine | Haloperidol | Loxapine | Lurasidone | Molindone | Olanzapine | Paliperidone | Placebo | Quetiapine | Risperidone | Trifluoperazine | Ziprasidone |
|---|---|---|---|---|---|---|---|---|---|---|---|---|---|---|---|
| Aripiprazole | NA | -0.174(-0.497,0.131) | 0.644(0.301,0.995) | -0.569(-1.027,-0.123) | -0.026(-0.32,0.277) | -0.313(-0.686,0.059) | -0.049(-0.352,0.227) | 0.144(-0.254,0.529) | 0.208(-0.058,0.475) | 0.001(-0.226,0.245) | -0.422(-0.609,-0.238) | -0.114(-0.37,0.141) | 0.155(-0.12,0.42) | -0.324(-0.736,0.094) | -0.262(-0.54,0.03) |
| Asenapine | 0.174(-0.131,0.497) | NA | 0.817(0.439,1.192) | -0.395(-0.889,0.112) | 0.147(-0.193,0.494) | -0.139(-0.553,0.264) | 0.125(-0.21,0.453) | 0.318(-0.095,0.739) | 0.382(0.076,0.712) | 0.175(-0.144,0.518) | -0.248(-0.485,0.007) | 0.06(-0.268,0.384) | 0.328(0.019,0.634) | -0.15(-0.599,0.291) | -0.089(-0.423,0.257) |
| Clozapine | -0.644(-0.995,-0.301) | -0.817(-1.192,-0.439) | NA | -1.213(-1.721,-0.701) | -0.67(-1.025,-0.299) | -0.957(-1.38,-0.529) | -0.693(-1.074,-0.31) | -0.5(-0.93,-0.059) | -0.436(-0.772,-0.101) | -0.643(-1.006,-0.281) | -1.066(-1.356,-0.766) | -0.758(-1.109,-0.423) | -0.489(-0.836,-0.153) | -0.967(-1.431,-0.496) | -0.906(-1.281,-0.524) |
| Fluphenazine | 0.569(0.123,1.027) | 0.395(-0.112,0.889) | 1.213(0.701,1.721) | NA | 0.543(0.111,0.993) | 0.256(-0.279,0.766) | 0.52(0.035,1) | 0.713(0.179,1.245) | 0.777(0.336,1.238) | 0.57(0.087,1.021) | 0.147(-0.273,0.559) | 0.455(-0.005,0.911) | 0.724(0.277,1.204) | 0.245(-0.31,0.764) | 0.306(-0.152,0.791) |
| Haloperidol | 0.026(-0.277,0.32) | -0.147(-0.494,0.193) | 0.67(0.299,1.025) | -0.543(-0.993,-0.111) | NA | -0.287(-0.66,0.065) | -0.023(-0.375,0.297) | 0.17(-0.235,0.595) | 0.234(-0.06,0.517) | 0.027(-0.305,0.346) | -0.396(-0.633,-0.159) | -0.088(-0.412,0.223) | 0.181(-0.141,0.493) | -0.297(-0.714,0.099) | -0.236(-0.582,0.087) |
| Loxapine | 0.313(-0.059,0.686) | 0.139(-0.264,0.553) | 0.957(0.529,1.38) | -0.256(-0.766,0.279) | 0.287(-0.065,0.66) | NA | 0.264(-0.153,0.649) | 0.457(-0.015,0.906) | 0.521(0.159,0.897) | 0.314(-0.058,0.685) | -0.109(-0.412,0.205) | 0.199(-0.174,0.587) | 0.468(0.089,0.844) | -0.011(-0.368,0.325) | 0.05(-0.344,0.441) |
| Lurasidone | 0.049(-0.227,0.352) | -0.125(-0.453,0.21) | 0.693(0.31,1.074) | -0.52(-1,-0.035) | 0.023(-0.297,0.375) | -0.264(-0.649,0.153) | NA | 0.193(-0.223,0.601) | 0.257(-0.045,0.575) | 0.05(-0.258,0.374) | -0.373(-0.584,-0.127) | -0.065(-0.384,0.245) | 0.203(-0.095,0.51) | -0.275(-0.691,0.166) | -0.214(-0.522,0.123) |
| Molindone | -0.144(-0.529,0.254) | -0.318(-0.739,0.095) | 0.5(0.059,0.93) | -0.713(-1.245,-0.179) | -0.17(-0.595,0.235) | -0.457(-0.906,0.015) | -0.193(-0.601,0.223) | NA | 0.064(-0.258,0.415) | -0.143(-0.536,0.27) | -0.566(-0.903,-0.221) | -0.258(-0.634,0.141) | 0.011(-0.337,0.382) | -0.468(-0.983,0.041) | -0.407(-0.797,0.027) |
| Olanzapine | -0.208(-0.475,0.058) | -0.382(-0.712,-0.076) | 0.436(0.101,0.772) | -0.777(-1.238,-0.336) | -0.234(-0.517,0.06) | -0.521(-0.897,-0.159) | -0.257(-0.575,0.045) | -0.064(-0.415,0.258) | NA | -0.207(-0.5,0.096) | -0.63(-0.83,-0.429) | -0.322(-0.614,-0.031) | -0.053(-0.306,0.191) | -0.532(-0.963,-0.144) | -0.471(-0.772,-0.16) |
| Paliperidone | -0.001(-0.245,0.226) | -0.175(-0.518,0.144) | 0.643(0.281,1.006) | -0.57(-1.021,-0.087) | -0.027(-0.346,0.305) | -0.314(-0.685,0.058) | -0.05(-0.374,0.258) | 0.143(-0.27,0.536) | 0.207(-0.096,0.5) | NA | -0.423(-0.639,-0.211) | -0.115(-0.409,0.17) | 0.154(-0.154,0.435) | -0.325(-0.77,0.1) | -0.264(-0.565,0.039) |
| Placebo | 0.422(0.238,0.609) | 0.248(-0.007,0.485) | 1.066(0.766,1.356) | -0.147(-0.559,0.273) | 0.396(0.159,0.633) | 0.109(-0.205,0.412) | 0.373(0.127,0.584) | 0.566(0.221,0.903) | 0.63(0.429,0.83) | 0.423(0.211,0.639) | NA | 0.308(0.096,0.519) | 0.577(0.363,0.78) | 0.098(-0.281,0.444) | 0.16(-0.069,0.385) |
| Quetiapine | 0.114(-0.141,0.37) | -0.06(-0.384,0.268) | 0.758(0.423,1.109) | -0.455(-0.911,0.005) | 0.088(-0.223,0.412) | -0.199(-0.587,0.174) | 0.065(-0.245,0.384) | 0.258(-0.141,0.634) | 0.322(0.031,0.614) | 0.115(-0.17,0.409) | -0.308(-0.519,-0.096) | NA | 0.269(-0.005,0.549) | -0.209(-0.635,0.214) | -0.148(-0.455,0.161) |
| Risperidone | -0.155(-0.42,0.12) | -0.328(-0.634,-0.019) | 0.489(0.153,0.836) | -0.724(-1.204,-0.277) | -0.181(-0.493,0.141) | -0.468(-0.844,-0.089) | -0.203(-0.51,0.095) | -0.011(-0.382,0.337) | 0.053(-0.191,0.306) | -0.154(-0.435,0.154) | -0.577(-0.78,-0.363) | -0.269(-0.549,0.005) | NA | -0.478(-0.897,-0.066) | -0.417(-0.72,-0.099) |
| Trifluoperazine | 0.324(-0.094,0.736) | 0.15(-0.291,0.599) | 0.967(0.496,1.431) | -0.245(-0.764,0.31) | 0.297(-0.099,0.714) | 0.011(-0.325,0.368) | 0.275(-0.166,0.691) | 0.468(-0.041,0.983) | 0.532(0.144,0.963) | 0.325(-0.1,0.77) | -0.098(-0.444,0.281) | 0.209(-0.214,0.635) | 0.478(0.066,0.897) | NA | 0.061(-0.363,0.484) |
| Ziprasidone | 0.262(-0.03,0.54) | 0.089(-0.257,0.423) | 0.906(0.524,1.281) | -0.306(-0.791,0.152) | 0.236(-0.087,0.582) | -0.05(-0.441,0.344) | 0.214(-0.123,0.522) | 0.407(-0.027,0.797) | 0.471(0.16,0.772) | 0.264(-0.039,0.565) | -0.16(-0.385,0.069) | 0.148(-0.161,0.455) | 0.417(0.099,0.72) | -0.061(-0.484,0.363) | NA |

**Appendix Table 15:** League table for all comparisons in CA network. Informative priors from GP studies using expert's opinion for $\beta$ and moderate downweight for GP studies evaluating interventions in $T_{\alpha} - T_{c}$.

| | Aripiprazole | Asenapine | Clozapine | Fluphenazine | Haloperidol | Loxapine | Lurasidone | Molindone | Olanzapine | Paliperidone | Placebo | Quetiapine | Risperidone | Trifluoperazine | Ziprasidone |
|---|---|---|---|---|---|---|---|---|---|---|---|---|---|---|---|
| Aripiprazole | NA | -0.176(-0.476,0.104) | 0.641(0.321,0.951) | -0.583(-1.02,-0.166) | -0.029(-0.297,0.256) | -0.266(-0.602,0.065) | -0.06(-0.335,0.199) | 0.115(-0.266,0.489) | 0.205(-0.049,0.464) | 0.005(-0.226,0.238) | -0.432(-0.608,-0.256) | -0.135(-0.37,0.1) | 0.148(-0.108,0.407) | -0.283(-0.66,0.08) | -0.264(-0.54,0.001) |
| Asenapine | 0.176(-0.104,0.476) | NA | 0.818(0.483,1.182) | -0.406(-0.873,0.06) | 0.147(-0.166,0.474) | -0.089(-0.462,0.27) | 0.116(-0.192,0.416) | 0.292(-0.105,0.674) | 0.382(0.104,0.671) | 0.182(-0.105,0.503) | -0.255(-0.474,-0.029) | 0.041(-0.253,0.341) | 0.324(0.027,0.617) | -0.107(-0.515,0.301) | -0.088(-0.382,0.209) |
| Clozapine | -0.641(-0.951,-0.321) | -0.818(-1.182,-0.483) | NA | -1.224(-1.694,-0.737) | -0.671(-0.996,-0.342) | -0.907(-1.293,-0.506) | -0.701(-1.031,-0.357) | -0.526(-0.956,-0.129) | -0.436(-0.741,-0.123) | -0.636(-0.98,-0.276) | -1.073(-1.338,-0.804) | -0.776(-1.101,-0.444) | -0.494(-0.82,-0.157) | -0.924(-1.369,-0.492) | -0.906(-1.24,-0.574) |
| Fluphenazine | 0.583(0.166,1.02) | 0.406(-0.06,0.873) | 1.224(0.737,1.694) | NA | 0.553(0.114,0.975) | 0.317(-0.18,0.789) | 0.523(0.059,0.962) | 0.698(0.199,1.221) | 0.788(0.358,1.23) | 0.588(0.157,1.041) | 0.151(-0.232,0.553) | 0.447(-0.021,0.899) | 0.73(0.294,1.181) | 0.3(-0.241,0.823) | 0.318(-0.134,0.775) |
| Haloperidol | 0.029(-0.256,0.297) | -0.147(-0.474,0.166) | 0.671(0.342,0.996) | -0.553(-0.975,-0.114) | NA | -0.236(-0.572,0.093) | -0.031(-0.348,0.26) | 0.145(-0.252,0.539) | 0.235(-0.048,0.513) | 0.035(-0.279,0.32) | -0.402(-0.637,-0.192) | -0.106(-0.418,0.175) | 0.177(-0.12,0.454) | -0.253(-0.64,0.132) | -0.235(-0.544,0.063) |
| Loxapine | 0.266(-0.065,0.602) | 0.089(-0.27,0.462) | 0.907(0.506,1.293) | -0.317(-0.789,0.18) | 0.236(-0.093,0.572) | NA | 0.205(-0.159,0.56) | 0.381(-0.044,0.814) | 0.471(0.139,0.819) | 0.271(-0.092,0.63) | -0.166(-0.453,0.125) | 0.13(-0.219,0.492) | 0.413(0.065,0.753) | -0.017(-0.337,0.304) | 0.001(-0.372,0.357) |
| Lurasidone | 0.06(-0.199,0.335) | -0.116(-0.416,0.192) | 0.701(0.357,1.031) | -0.523(-0.962,-0.059) | 0.031(-0.26,0.348) | -0.205(-0.56,0.159) | NA | 0.175(-0.194,0.568) | 0.265(-0.003,0.553) | 0.066(-0.238,0.366) | -0.372(-0.574,-0.162) | -0.075(-0.347,0.205) | 0.208(-0.061,0.486) | -0.223(-0.617,0.175) | -0.204(-0.483,0.086) |
| Molindone | -0.115(-0.489,0.266) | -0.292(-0.674,0.105) | 0.526(0.129,0.956) | -0.698(-1.221,-0.199) | -0.145(-0.539,0.252) | -0.381(-0.814,0.044) | -0.175(-0.568,0.194) | NA | 0.09(-0.245,0.441) | -0.11(-0.497,0.289) | -0.547(-0.87,-0.212) | -0.25(-0.62,0.131) | 0.033(-0.324,0.378) | -0.398(-0.849,0.065) | -0.379(-0.771,0.008) |
| Olanzapine | -0.205(-0.464,0.049) | -0.382(-0.671,-0.104) | 0.436(0.123,0.741) | -0.788(-1.23,-0.358) | -0.235(-0.513,0.048) | -0.471(-0.819,-0.139) | -0.265(-0.553,0.003) | -0.09(-0.441,0.245) | NA | -0.2(-0.483,0.081) | -0.637(-0.827,-0.457) | -0.341(-0.6,-0.071) | -0.058(-0.302,0.176) | -0.488(-0.869,-0.111) | -0.47(-0.74,-0.204) |
| Paliperidone | -0.005(-0.238,0.226) | -0.182(-0.503,0.105) | 0.636(0.276,0.98) | -0.588(-1.041,-0.157) | -0.035(-0.32,0.279) | -0.271(-0.63,0.092) | -0.066(-0.366,0.238) | 0.11(-0.289,0.497) | 0.2(-0.081,0.483) | NA | -0.437(-0.643,-0.231) | -0.141(-0.423,0.124) | 0.142(-0.131,0.423) | -0.288(-0.676,0.103) | -0.27(-0.569,0.011) |
| Placebo | 0.432(0.256,0.608) | 0.255(0.029,0.474) | 1.073(0.804,1.338) | -0.151(-0.553,0.232) | 0.402(0.192,0.637) | 0.166(-0.125,0.453) | 0.372(0.162,0.574) | 0.547(0.212,0.87) | 0.637(0.457,0.827) | 0.437(0.231,0.643) | NA | 0.297(0.098,0.499) | 0.58(0.395,0.774) | 0.149(-0.188,0.485) | 0.168(-0.045,0.377) |
| Quetiapine | 0.135(-0.1,0.37) | -0.041(-0.341,0.253) | 0.776(0.444,1.101) | -0.447(-0.899,0.021) | 0.106(-0.175,0.418) | -0.13(-0.492,0.219) | 0.075(-0.205,0.347) | 0.25(-0.131,0.62) | 0.341(0.071,0.6) | 0.141(-0.124,0.423) | -0.297(-0.499,-0.098) | NA | 0.283(0.027,0.544) | -0.148(-0.533,0.241) | -0.129(-0.432,0.158) |
| Risperidone | -0.148(-0.407,0.108) | -0.324(-0.617,-0.027) | 0.494(0.157,0.82) | -0.73(-1.181,-0.294) | -0.177(-0.454,0.12) | -0.413(-0.753,-0.065) | -0.208(-0.486,0.061) | -0.033(-0.378,0.324) | 0.058(-0.176,0.302) | -0.142(-0.423,0.131) | -0.58(-0.774,-0.395) | -0.283(-0.544,-0.027) | NA | -0.431(-0.803,-0.045) | -0.412(-0.693,-0.13) |
| Trifluoperazine | 0.283(-0.08,0.66) | 0.107(-0.301,0.515) | 0.924(0.492,1.369) | -0.3(-0.823,0.241) | 0.253(-0.132,0.64) | 0.017(-0.304,0.337) | 0.223(-0.175,0.617) | 0.398(-0.065,0.849) | 0.488(0.111,0.869) | 0.288(-0.103,0.676) | -0.149(-0.485,0.188) | 0.148(-0.241,0.533) | 0.431(0.045,0.803) | NA | 0.019(-0.377,0.421) |
| Ziprasidone | 0.264(-0.001,0.54) | 0.088(-0.209,0.382) | 0.906(0.574,1.24) | -0.318(-0.775,0.134) | 0.235(-0.063,0.544) | -0.001(-0.357,0.372) | 0.204(-0.086,0.483) | 0.379(-0.008,0.771) | 0.47(0.204,0.74) | 0.27(-0.011,0.569) | -0.168(-0.377,0.045) | 0.129(-0.158,0.432) | 0.412(0.13,0.693) | -0.019(-0.421,0.377) | NA |

**Appendix Table 16:** League table for all comparisons in CA network using a NMA model with non-informative priors.

| | Aripiprazole | Asenapine | Clozapine | Fluphenazine | Haloperidol | Loxapine | Lurasidone | Molindone | Olanzapine | Paliperidone | Placebo | Quetiapine | Risperidone | Trifluoperazine | Ziprasidone |
|---|---|---|---|---|---|---|---|---|---|---|---|---|---|---|---|
| Aripiprazole | NA | -0.046(-0.517,0.44) | 0.628(-0.142,1.404) | -1.555(-2.581,-0.524) | -0.193(-0.927,0.525) | -0.295(-1.284,0.633) | 0.05(-0.388,0.512) | 0.361(-0.212,0.97) | 0.317(-0.149,0.769) | -0.036(-0.342,0.282) | -0.433(-0.718,-0.137) | -0.044(-0.381,0.293) | 0.24(-0.2,0.66) | -0.259(-1.383,0.801) | -0.281(-0.744,0.224) |
| Asenapine | 0.046(-0.44,0.517) | NA | 0.674(-0.084,1.466) | -1.509(-2.529,-0.474) | -0.147(-0.916,0.621) | -0.249(-1.254,0.7) | 0.096(-0.434,0.608) | 0.407(-0.234,1.044) | 0.362(-0.16,0.863) | 0.01(-0.477,0.519) | -0.387(-0.762,-0.011) | 0.001(-0.471,0.515) | 0.286(-0.262,0.763) | -0.213(-1.321,0.908) | -0.235(-0.756,0.294) |
| Clozapine | -0.628(-1.404,0.142) | -0.674(-1.466,0.084) | NA | -2.183(-3.109,-1.171) | -0.821(-1.519,-0.16) | -0.923(-1.823,-0.014) | -0.578(-1.352,0.227) | -0.267(-1.009,0.509) | -0.312(-0.917,0.308) | -0.664(-1.439,0.154) | -1.061(-1.731,-0.351) | -0.673(-1.415,0.09) | -0.388(-1.072,0.31) | -0.887(-1.869,0.091) | -0.909(-1.721,-0.094) |
| Fluphenazine | 1.555(0.524,2.581) | 1.509(0.474,2.529) | 2.183(1.171,3.109) | NA | 1.362(0.642,2.089) | 1.26(0.329,2.21) | 1.604(0.555,2.706) | 1.916(0.877,2.953) | 1.871(0.958,2.833) | 1.518(0.502,2.58) | 1.122(0.12,2.152) | 1.51(0.51,2.561) | 1.795(0.828,2.733) | 1.296(0.265,2.343) | 1.274(0.223,2.373) |
| Haloperidol | 0.193(-0.525,0.927) | 0.147(-0.621,0.916) | 0.821(0.16,1.519) | -1.362(-2.089,-0.642) | NA | -0.102(-0.718,0.535) | 0.243(-0.507,1.026) | 0.554(-0.186,1.301) | 0.509(-0.046,1.115) | 0.157(-0.605,0.938) | -0.24(-0.892,0.442) | 0.148(-0.577,0.902) | 0.433(-0.191,1.078) | -0.066(-0.812,0.705) | -0.088(-0.863,0.684) |
| Loxapine | 0.295(-0.633,1.284) | 0.249(-0.7,1.254) | 0.923(0.014,1.823) | -1.26(-2.21,-0.329) | 0.102(-0.535,0.718) | NA | 0.345(-0.618,1.372) | 0.656(-0.285,1.616) | 0.611(-0.203,1.455) | 0.259(-0.702,1.243) | -0.138(-1.033,0.834) | 0.251(-0.684,1.232) | 0.535(-0.337,1.448) | 0.036(-0.413,0.494) | 0.014(-0.953,1.018) |
| Lurasidone | -0.05(-0.512,0.388) | -0.096(-0.608,0.434) | 0.578(-0.227,1.352) | -1.604(-2.706,-0.555) | -0.243(-1.026,0.507) | -0.345(-1.372,0.618) | NA | 0.311(-0.324,0.948) | 0.267(-0.28,0.78) | -0.086(-0.579,0.426) | -0.482(-0.85,-0.112) | -0.094(-0.58,0.393) | 0.19(-0.358,0.677) | -0.309(-1.437,0.779) | -0.331(-0.852,0.186) |
| Molindone | -0.361(-0.97,0.212) | -0.407(-1.044,0.234) | 0.267(-0.509,1.009) | -1.916(-2.953,-0.877) | -0.554(-1.301,0.186) | -0.656(-1.616,0.285) | -0.311(-0.948,0.324) | NA | -0.045(-0.509,0.41) | -0.397(-0.992,0.199) | -0.794(-1.316,-0.258) | -0.406(-0.985,0.196) | -0.121(-0.602,0.327) | -0.62(-1.655,0.444) | -0.642(-1.265,0.037) |
| Olanzapine | -0.317(-0.769,0.149) | -0.362(-0.863,0.16) | 0.312(-0.308,0.917) | -1.871(-2.833,-0.958) | -0.509(-1.115,0.046) | -0.611(-1.455,0.203) | -0.267(-0.78,0.28) | 0.045(-0.41,0.509) | NA | -0.353(-0.856,0.176) | -0.749(-1.119,-0.342) | -0.361(-0.819,0.119) | -0.076(-0.431,0.285) | -0.575(-1.516,0.392) | -0.597(-1.127,-0.04) |
| Paliperidone | 0.036(-0.282,0.342) | -0.01(-0.519,0.477) | 0.664(-0.154,1.439) | -1.518(-2.58,-0.502) | -0.157(-0.938,0.605) | -0.259(-1.243,0.702) | 0.086(-0.426,0.579) | 0.397(-0.199,0.992) | 0.353(-0.176,0.856) | NA | -0.396(-0.757,-0.066) | -0.008(-0.461,0.406) | 0.276(-0.213,0.738) | -0.223(-1.302,0.872) | -0.245(-0.764,0.289) |
| Placebo | 0.433(0.137,0.718) | 0.387(0.011,0.762) | 1.061(0.351,1.731) | -1.122(-2.152,-0.12) | 0.24(-0.442,0.892) | 0.138(-0.834,1.033) | 0.482(0.112,0.85) | 0.794(0.258,1.316) | 0.749(0.342,1.119) | 0.396(0.066,0.757) | NA | 0.388(0.08,0.705) | 0.673(0.305,1.001) | 0.174(-0.868,1.224) | 0.152(-0.226,0.536) |
| Quetiapine | 0.044(-0.293,0.381) | -0.001(-0.515,0.471) | 0.673(-0.09,1.415) | -1.51(-2.561,-0.51) | -0.148(-0.902,0.577) | -0.251(-1.232,0.684) | 0.094(-0.393,0.58) | 0.406(-0.196,0.985) | 0.361(-0.119,0.819) | 0.008(-0.406,0.461) | -0.388(-0.705,-0.08) | NA | 0.285(-0.162,0.696) | -0.214(-1.297,0.857) | -0.236(-0.718,0.288) |
| Risperidone | -0.24(-0.66,0.2) | -0.286(-0.763,0.262) | 0.388(-0.31,1.072) | -1.795(-2.733,-0.828) | -0.433(-1.078,0.191) | -0.535(-1.448,0.337) | -0.19(-0.677,0.358) | 0.121(-0.327,0.602) | 0.076(-0.285,0.431) | -0.276(-0.738,0.213) | -0.673(-1.001,-0.305) | -0.285(-0.696,0.162) | NA | -0.499(-1.479,0.518) | -0.521(-1.011,0.009) |
| Trifluoperazine | 0.259(-0.801,1.383) | 0.213(-0.908,1.321) | 0.887(-0.091,1.869) | -1.296(-2.343,-0.265) | 0.066(-0.705,0.812) | -0.036(-0.494,0.413) | 0.309(-0.779,1.437) | 0.62(-0.444,1.655) | 0.575(-0.392,1.516) | 0.223(-0.872,1.302) | -0.174(-1.224,0.868) | 0.214(-0.857,1.297) | 0.499(-0.518,1.479) | NA | -0.022(-1.114,1.079) |
| Ziprasidone | 0.281(-0.224,0.744) | 0.235(-0.294,0.756) | 0.909(0.094,1.721) | -1.274(-2.373,-0.223) | 0.088(-0.684,0.863) | -0.014(-1.018,0.953) | 0.331(-0.186,0.852) | 0.642(-0.037,1.265) | 0.597(0.04,1.127) | 0.245(-0.289,0.764) | -0.152(-0.536,0.226) | 0.236(-0.288,0.718) | 0.521(-0.009,1.011) | 0.022(-1.079,1.114) | NA |